
\magnification=1200
\baselineskip=16pt

\font\caps=cmssbx10 at 12truept
\font\ccaps=cmssbx10 at 16truept
 at 12truept

\def\lot{{\;\buildrel < \over \sim \;}}
\def\got{{\;\buildrel > \over \sim \;}}

\def\half{\hbox{$\textstyle {1\over 2}$}}
\def\etal{{\it et al.}}

\def\la{\hbox{$\langle$}}
\def\ra{\hbox{$\rangle$}}

\def\va{{\vec a}}   
 \def\vf{{\vec f}}  
  \def\vk{{\vec k}} 
   
\def\vq{{\vec q}} \def\vr{{\vec r}}  
\def\vu{{\vec u}} \def\vv{{\vec v}}  \def\vx{{\vec x}}

 \def\vB{{\vec B}}  
 \def\vF{{\vec F}} \def\vG{{\vec G}} \def\vH{{\vec H}}

 \def\vR{{\vec R}}

\def\CD{\hbox{$\cal D$}}  
 \def\CH{\hbox{$\cal H$}} 
  
  \def\CO{\hbox{$\cal O$}}

 \def\CZ{\hbox{$\cal Z$}}

\def\pmb#1{\setbox0=\hbox{#1}
  \kern-.025em\copy0\kern-\wd0
  \kern.05em\copy0\kern-\wd0
  \kern-.025em\raise.0433em\box0 }

\def\eq{_{{\rm eq}}}
\def\ep{\epsilon}
\def\eps{\varepsilon}
\def\te{\tilde\varepsilon_d}
\def\tel{\tilde\varepsilon_1}

\def\vnab{{\pmb{$\nabla$}}}

\def\'{^{\,\prime}}
\def\u{{\vec u}}
\def\r{{\vec r}}
\def\B{{\vec B}}
\def\0{{\vec 0}}

\def\H{{\vec H}}
\def\G{{\vec G}}
\def\c{{\vec c}}


\centerline{\ccaps Interstitials, Vacancies and Supersolid Order in Vortex
Crystals}
\bigskip
\centerline{by}
\bigskip
\centerline{Erwin Frey}
\centerline{Institut f\"ur Theoretische Physik}
\centerline{Physik Department der Technischen Universit\"at M\"unchen}
\centerline{D-85747 Garching, Germany}
\bigskip
\centerline{David R.~Nelson  and Daniel S. Fisher}
\centerline{Lyman Laboratory of Physics}
\centerline{Harvard University}
\centerline{Cambridge, Massachusetts \ 02138}
\vskip 1.5truein
\centerline{\caps Abstract}

Interstitials and vacancies in the Abrikosov phase of clean
Type~II superconductors
are line imperfections, which cannot extend across macroscopic equilibrated
samples at low
temperatures. We argue that the entropy associated with line wandering
nevertheless can cause
these defects to proliferate at a sharp transition which will
exist if this occurs below
the temperature at which the crystal actually melts. Vortices are both
entangled and
crystalline in the resulting ``supersolid'' phase, which
in a dual ``boson'' analog system
is closely related to a
two-dimensional quantum crystal of He$^4$ with interstitials or vacancies in
its ground state.
The supersolid  {\it must} occur for $B\gg B_\times$, where $B_\times$ is the
decoupling field above which vortices begin to behave two-dimensionally.
Numerical calculations show that interstitials, rather than vacancies, are the
preferred defect for $B\gg \phi_0/\lambda_\perp^2$,
and allow us to estimate whether
proliferation also occurs for $B\,\lot\,B_\times$.
The implications of the supersolid
phase for transport measurements, dislocation configurations and neutron
diffraction are discussed.
\vfill\eject

\baselineskip=16pt plus 1pt minus 1pt
\parskip=10pt plus 1pt minus 1pt
\noindent
{\caps 1. INTRODUCTION }

Fluctuations, especially in high-temperature superconductors, play a prominent
role in determining vortex configurations in Type~II materials in an external
field [1]. It now appears, for example, that clean
single crystal samples of YBCO (in the absence of twin boundary pinning)
melt via a first-order phase transition [2]
at a temperature $T_m$ well below the upper critical field line H$_{c2}(T)$
predicted by mean field theory [3--5]. Point disorder, in the form of
oxygen vacancies, does not seem to affect this phase transition
strongly in YBCO. It is quite
possible that the disorder-induced translational correlation length [6] $R_a$
greatly exceeds the vortex spacing for $T<T_m$ in the field range for which
the transition is first order.
Although in the presence of randomness
the low temperature phase is not, strictly speaking, a solid,
the thermodynamic properties at and near the phase transition should be similar
to
those in the absence of randomness.

Consider the thermal excitations about the crystalline state on
scales less than $R_a$. We assume for simplicity vortices which are
perpendicular on average to the CuO$_2$ planes, i.e.\ parallel to the
$z$-direction.  The finite reduction
of the Debye-Waller factor associated with the translational
order parameter
$$
\eqalignno{
\rho_{\G}(T)&= \langle e^{i\G\cdot\u(\r_\perp,z)}
\rangle\cr
&=\exp\left[-\half\; G_iG_j\langle
u_i(\r_\perp,z)u_j(\r_\perp,z)\rangle\right]
&(1.1)\cr}
$$
by {\it phonons} is discussed in Ref.~7. Here, $\G$ is a reciprocal lattice
vector and
$\u(\r_\perp,z)$ is the displacement field of a flux lattice with
vortices parallel on average to the $z$ direction. {\it Dislocation loops}
are a topologically distinct excitation which, when they proliferate
at a melting transition, drive $\rho_{\G}(T)$ to zero and
can lead to a hexatic flux liquid with residual
bond orientation order [8]. Isolated dislocation loops are far more
constrained than their counterparts in crystals of point particles:
dislocation loops in fact must lie in a plane spanned by their Burgers
vector and the average field direction [9]---see Fig.~1a.

Vacancies and interstitials differ even more dramatically from the analogous
defects
in crystals of point particles. The number of flux lines is conserved, which
means
that these defects are {\it lines} instead of points. The point-like nature of
vacancies and interstitials in conventional crystals ensures that they are
present in equilibrium at all finite temperatures for entropic reasons [10].
However,
because such imperfections have an energy proportional to their {\it length}
in flux crystals, they cannot extend completely across an
equilibrated macroscopic sample at low temperatures. A typical
fluctuation at low temperatures might consist of the vacancy-interstitial
pair shown in Fig.~1b. Unlike the dislocation loop in Fig.~1a,
this loop is not constrained to lie in a single plane.
This configuration also provides a mechanism by which vortices near the loop
may jog one lattice constant to the right as $z$ varies.

Defect loops of this type can have important dynamical consequences.
In two dimensions, point-like vacancies and interstitials probably dominate
the resistive properties of a weakly pinned lattice in a superconducting
film [13, 14]. The same may well be true of vacancy-interstitial loops
and lines in three dimensions as discussed in Section~5.

Although the energy of vacancy and interstitial lines is proportional
to their length, it is nevertheless possible for these defects to
``proliferate'' (i.e., to become infinitely long) at high
temperatures for entropic reasons. Consider, for example, a vacancy
wandering across a macroscopic sample of thickness $L$, as in Fig.~2.
To estimate the free energy of this defect, we describe its
trajectory along the $z$ axis by a function $\r_d(z)$ and write its
partition function as a functional integral,
$$
\CZ_d=e^{-\epsilon_d L/T}
\int\CD\r_d(z)\exp
\left\{-{1\over T}\int_0^L
dz\left[{\half}\te
\left({d\r_d\over dz}\right)^2
+U_\ell(\r_d)\right]\right\}\;.
\eqno(1.2)
$$
Here $\eps_d$ is the energy of an isolated defect and $\te$ is its tilt
energy, defined in analogy with similar quantities for isolated flux
lines near H$_{ c1}$ [7] and $U_\ell(\r_d)$ is a periodic lattice potential
with minima at the sites of the triangular Abrikosov crystal. Implicit in the
path integral (1.2) is a length scale $\ell_z$ which is the average distance
along $z$ between hops of the vacancy from one lattice position to another.
As a crude estimate of the path integral we replace it by
$\exp(-\eps_d L/T)$ $6^{L/\ell_z}$, since the vacancy has six directions
in which to hop on a triangular lattice. The free energy $F_d=-T\ln Z_d$ is
thus
$$
F_d\approx \eps_dL-{T\over\ell_z}\;L\ln 6\;,
\eqno(1.3{\rm a})
$$
which becomes negative for $T>T_d $, where
$$
T_d=\eps_d\ell_z/\ln 6\;.
\eqno(1.3{\rm b})
$$
Above this temperature (provided the crystal does not melt first),
vacancies (or interstitials) will proliferate in a crystalline phase.
If these defects do not become strongly pinned by point disorder,
one might then expect
a {\it linear} contribution to the resistivity due to defect
motion within pinned bulk crystallites, and a vortex glass transition
[12] for weak pinning might then  occur near $T_d$.
More precise  estimates of the
proliferation temperature will be represented later in this paper. A related
phenomenon was suggested by Feigel'man \etal~[13], who predicted the
unbinding of ``quarters'' (i.e., quartets) of dislocations above the
decoupling field $B_\times$ in highly anisotropic superconductors (see below).

The phase in which defects such as interstitials and vacancies proliferate is
in fact both crystalline and entangled. Regarded in light of the analogy
between
thermally excited flux lines and two-dimensional bosons [15], it represents
a ``supersolid'' quantum crystal, in which vacancies and interstitials are
incorporated into the ground state [15--20]. The possibility of a supersolid
phase for flux lines in Type II superconductors was first noted on the basis
of the boson analogy by M.P.A.~Fisher and Lee [21]. Somewhat paradoxically,
the ``supersolid'' is actually {\it less} superconducting for the BCS
condensate electrons than a conventional vortex lattice phase.
An alternative name is an ``incommensurate solid'' or a
``vortex density wave.''

Figure 3 illustrates the connection between defect proliferation and the boson
order parameter for flux lines [22]. This figure represents a low-temperature
contribution to the order parameter correlation function
$$
G(\r_\perp,\r_\perp\';z,z')=
\langle\psi(\r_\perp,z)\psi^*(\r_\perp\',z')\rangle\;,
\eqno(1.4)
$$
where $\psi(\r_\perp,z)$ and $\psi^*(\r_\perp\',z')$ are destruction and
creation operators for {\it flux lines} moving along the $z$ axis,
which plays the role of ``time'' for the ``bosons.'' The
composite operator in Eq. (1.4) creates an extra line at $(\r_\perp\',z')$ and
destroys an existing line at $(\r_\perp,z)$. The lowest energy configuration
is then a line of vacancies (for $z'>z$, as in Fig.~3) or interstitials (for
$z'<z)$ connecting $(\r_\perp\',z')$ to $(\r_\perp,z)$. At low temperatures,
the defect line joining the head to the tail costs an energy of order
$f_d(\hat s)s$,
where $s=\sqrt{(\r_\perp-\r_\perp\')^2+(z-z')^2}$ is the length
of the string in direction $\hat s$ and $f_d(\hat s)$ is the angle-dependent
defect-free energy per unit length. The
correlation function (1.4) then decays exponentially to zero, like
$\exp(-f_ds/T)$, as $|\r_\perp-\r_\perp\'|\rightarrow\infty$ with $z$
and $z'$ fixed. Above $T_d$, $f_d$ changes sign, defects proliferate,
and there is long-range order in $G(\r_\perp,\r_\perp\';z,z')$,
$$
\eqalignno{
\lim_{|\r_\perp-\r_\perp\'|\rightarrow\infty}
G(\r_\perp,\r_\perp\';z,z')&=
\langle\psi(\r_\perp,z)\rangle
\langle\psi^*(\r_\perp\',z')\rangle\cr
&\not= 0\quad.
&(1.5)\cr}
$$

The physical meaning of Eq. (1.5) is that the free energy of the extra line
segment from
$(\r,z)$ to $(\r',z')$ in the limit of large
separations remains {\it finite}. The quantity  $-\ln G$ measures
the free energy of a monopole anti-monopole pair at a $(\r,z)$ and
$(\r',z')$ respectively. In the conventional solid phase the magnetic
monopoles are confined by a linear potential, while when the defects
proliferate, the monopoles are unconfined as in a normal metal
with finite total free energy cost.
Entanglement of vortex lines in the crystal will be catalyzed by the
proliferation of vacancies and interstitials, since these allow fluxons
to easily move perpendicular to the $z$ axis. Of course, entanglement
will also arise if the crystal melts directly into a flux liquid. In
either case the boson order parameter becomes nonzero,
$$
\psi_0(T)=\langle\psi(\r_\perp,z)\rangle\not= 0\quad.
\eqno(1.6)
$$

Note that Eq. (1.6) gives a precise meaning to the concept of
``entanglement'' as used here and elsewhere to describe ``superfluid''
phases within the boson analogy. Strictly speaking, the concept
of ``tangled vortex lines'' does not distinguish sharply between
low and high temperature phases
for lines in an infinitely thick sample.
Indeed, in  {\it all} phases discussed here---solid, ``supersolid'' and
fluid---a
vortex line will wander as a function of $z$ as a random walker on sufficiently
large
scales. Although this wandering will be more pronounced in the higher
temperature
phases, it does  not uniquely  distinguish them from the
solid. The presence or absence of particle diffusion does not sharply
distinguish
solids from liquids of point particles for similar reasons. Note also that
labelling of vortices becomes ambiguous when the lines pass within a
coherence length of each other---one must sum over ``direct'' and
``exchange'' connection possibilities to define the statistical mechanics.
Entanglement also has a quantitiative meaning for
{\it dynamics}: if lines  can only cross or recombine by overcoming
large free energy barriers, such lines  are {\it dynamically} entangled,
similar to a dislocation tangle in a work hardened metal.
In this paper we will continue the usage in the recent literature of
referring only to phases which are ``superfluid'' in the sense of Eq.~(1.6) as
entangled.

Figure 4 illustrates two distinct scenarios for
phase diagrams of vortices with increasing
temperature, which we call ``Type I'' and ``Type II'' melting. In
Type~I melting a first-order transition separates a line crystal with
$\rho_{\G}\not=0$ from a flux liquid with $\psi_0\not= 0$
which may or may not be hexatic. In Type~II
melting, {\it both} order parameters are nonzero in an intermediate
``supersolid'' phase. As discussed in Sec.~4,
vacancies or interstitials enter the Abrikosov flux
lattice at $T_d$ in much the same way as the flux lines penetrate the
Meissner phase at H$_{ c1}$. Although this transition can in principle be
continuous, even in the presence of strong thermal fluctuations [17], the
melting of the supersolid into a liquid is likely to remain a first-order
transition, since, as far as is known, continuous melting transitions
in which materials lose crystallinity in two or more directions
do not occur in three dimensions.

Can a real supersolid phase exist in quantum crystals?
According to the review by Andreev [23], ``the experimental data available
at present show that the possibility ...(of a supersolid)... can hardly
take place in [bulk] $^4$He crystals.''
In thin films of $^4$He, supersolid phases {\it can}, however, exist. On
substrates in which one or two layers of incommensurate solid $^4$He form,
there
can be a regime in which the atoms in the next partially filled
layer are superfluid; this is a 2D supersolid phase even though there
are almost no vacancies or interstitials in the close-packed solid
layers.
In 2D electron crystals at zero temperature, some kind of solid phase
in which interstitials proliferate may  occur [26b]. However,
since electrons are fermions, such a phase would probably not be
supersolid, expect perhaps at extremely low temperatures.

In the case of flux lines, the softer (longitudinal) nature of the
interaction between lines makes a supersolid less unlikely than in bulk $^4$He.
However, in  the limit of {\it continuous} flux lines,
where the boson analogy is best, we shall see that such a phase
is still improbable.
The new ingredient in the flux system, which does {\it not} have an analog
for bosons, is the discreteness of the layers, i.e. discreteness in the
time-like direction. As we shall see, this will
definintely cause a supersolid for sufficiently large $B$ in strongly
anisotropic layered superconductors.
The characteristic magnetic field above which the layering  becomes
important is the crossover field [12, 24]
$$
B_\times \sim\gamma^2\,{\phi_0\over d^2_0}\;,
\eqno(1.6)
$$
where $d_0$ is the layer spacing and
$$
\gamma^2\equiv{M_\perp\over M_z}
\eqno(1.7)
$$
the effective mass anisotropy.
Physically, this field represents the crossover between different behaviors for
the
energy at a ``jog'' where one line is shifted by an intervortex spacing
$a_0=(\phi_0/B)^{1/2}$ from one layer to the next. A jog with shift  distance
$a<d_0/\gamma$  costs
an energy quadratic in $a$, while for
$a>d_0/\gamma$, the energy cost will be linear in $a$ and a jog will
tend to spread out over more than one layer. For jogs with shifts $a_0$, the
former will obtain for $B\gg B_\times$.
In a sufficiently anisotropic system, $B_\times\ll H_{c2}(T=0)$ and
a regime exists, $B\gg B_\times$, in which the vortex fluctuations
are predominantly two dimensional.

Figure 5 summarizes our conclusions
about the phase diagram as function of magnetic
field $B$ (i.e., vortex density) and temperature $T$, with external field $\vec
H$ is
parallel to the $\hat c$ axis. In a sufficiently anisotropic system,
the supersolid {\it must} exist for $B\gg B_\times$,
where vortices in different copper-oxide layers are approximately decoupled
by thermal
fluctuations [12,24]. The defects in this regime are equivalent
to the ``unbound
dislocation quartets'' discussed by Feigel'man \etal~[13]. The existence of
a supersolid for fields $B\,\lot\,B_\times$
is a more delicate matter, which we shall
discuss more quantitatively in Sec.~4.  For fields $B\gg B_{ c1}\equiv
\phi_0/\lambda_\perp^2$, so that vortices interact
via a logarithmic potential, the
preferred defects are interstitials instead of vacancies. Similar conclusions
were reached by Fisher \etal \ and by Cockayne and Elser
for electrons interacting with a $1/r$ potential
in two dimensions [25]. At low fields $(B\;\lot\;B_{ c1})$,
we expect that vacancies are the
favored defect, as in most crystals with short-range interactions.
Although a supersolid is possible in principle for any field $B<B_\times$, the
numerical estimates of Sec.~4 indicate that $B\;\got\ B_\times$
is required for this new phase to exist in high-T$_{\rm c}$ superconductors.

Even if an {\it equilibrium} supersolid does not exist for $B<B_\times$,
vacancies and interstitials may still appear for nonequilibrium reasons. If,
for example, vortices pass through a first-order freezing transition upon
cooling under conditions of constant $B$, the system will initially phase
separate into crystallites coexisting with a liquid of a different density.
When the liquid disappears at a lower temperature, the crystal density
must change to keep the macroscopic $B$ field constant. In the presence
of pinning, kinetic constraints may force the crystal to absorb interstitials
or vacancies rather than to changing its overall lattice constant. If
freezing occurs {\it above} the field corresponding to maximum melting
temperature in Fig.~5, a nonequilibrium concentration of interstitials
would then result [26].

In Section 2 we derive simple estimates for the defect unbinding temperature in
various regimes, and compare these to the melting temperature. Numerical
calculations of the energies of various kinds of straight defect lines in the
high field regime are presented in Section~3. In Section~4 these energies are
used
to estimate $T_d$ more quantitatively for $B_{ c1}<\,<B\;\lot\;B_\times$.
We also discuss the consequences of supersolid order for neutron
diffraction and theories of the melting transition. And, finally
the implications of a supersolid phase for resistivity
measurements, neutron diffraction and dislocation configurations.
\vfill\eject
\noindent
{\caps 2. ESTIMATES OF DEFECT UNBINDING TRANSITIONS}

\noindent
{\caps 2.1 Model Parameters and Field Regimes}

To obtain the important physical parameters in a thermally excited vortex
lattice, it is instructive to
examine one representative fluxon in the confining
potential provided by its surrounding vortices in a triangular lattice.
$$
\CZ_1(\r_\perp,\0;L)=
\int_{\r(0)=\0}^{\r(L)=\r_\perp}
\CD\r(z)\exp\left\{-{1\over T}\int_0^L
\left[\half\tel\left({d\r(z)\over dz}\right)^2+V_1[\r(z)]\right]dz\right\}.
\eqno(2.1)
$$
As we shall see, this simple model gives predictions for melting equivalent
to those obtained via ``nonlocal'' elastic constants and the Lindemann
criterion [27]. Closely related models (such as Eq.~(1.2)) will be used to
determine when defects proliferate. The effective potential
$V_1[\r(z)]$ in (2.1)
arises from a microscopic pairwise interaction $V[\r_{ij}(z)]$
between two parallel flux lines $\r_i(z)$ and $\r_j(z)$ with
separation $\r_{ij}(z)$. In the London approximation, this
potential is~[1]
$$
V(r_{ij})=2\eps_0K_0(r_{ij}/\lambda_\perp)
\eqno(2.2)
$$
where
$$
\eps_0=
(\phi_0/4\pi\lambda_\perp)^2\;,
\eqno(2.3)
$$
with $\lambda_\perp$ the in-plane London penetration depth is the
energy scale per unit length.  The Bessel
function $K_0(x)\approx\ln(x)$ for $x\ll 1$, and $K_0(x)
\approx\left({\pi\over 2x}\right)^{1/2}e^{-x}$ for $x$ large. The
parameter $\tel$ is the tilt energy of a flux line, given
approximately by
$$
\tel\approx \gamma^2
\eps_0\ln(a_0/\xi_\perp)\qquad\qquad \qquad
(B\gg B_{ c1})
\eqno(2.4{\rm a})
$$
where $\gamma^2\equiv M_\perp/M_z\ll 1$ is the mass anisotropy,
$a_0=(\phi_0/B)^{1/2}$, and
$\xi_\perp$ is the in-plane coherence length. Equation~(2.4a)
follows from the wave-vector-dependent
tilt modulus evaluated in the short distance
regime relevant for melting [24]. This result applies only when
$B\gg B_{c1}\equiv\phi_0/\lambda_\perp^2$, so that we may neglect the
electromagnetic coupling between CuO$_2$ planes. When $B\;\lot\;B_{c1}$, this
coupling is important and the tilt energy becomes [12]
$$
\tel\approx\eps_0\;,\qquad\qquad\qquad  (B\;\lot\;B_{ c1})\;.
\eqno(2.4{\rm b})
$$

Although the tilt modulus has more generally a complicated wavevector
dependence
[28], it is approximately constant over the (short distance) length scales of
interest to us here. ``Nonlocal'' (in $z$) contributions to the interaction
potential [29] are similarly unimportant provided [30]
$$
\langle\left|{d\r\over dz}\right|^2\rangle\ll\gamma^{-2}\;,
\eqno(2.5)
$$
which is well satisfied throughout the crystalline phase. As discussed
by Brandt [31], the same criterion justifies the neglect of higher order terms
in $[d\r(z)/dz]^2$ in Eq.~(2.1).

Three important field regimes for fluctuations in vortex crystals are easily
extracted with this approach.
We first rewrite this imaginary time path integral
as a quantum mechanical matrix element [32],
$$
\CZ(\r_\perp,\0;L)=\langle\r_\perp|e^
{-L\CH/T}|\0\rangle
\eqno(2.6)
$$
where $|\0\rangle$ is an initial state localized at $\0$, $\langle\r_\perp|$ is
a final state localized at $\r_\perp$, and the ``Hamiltonian'' $\CH$ is
$$
\CH=-{T^2\over 2\tel}\;
\nabla_\perp^2+V_1(\r_\perp)\;.
\eqno(2.7)
$$
The probability of finding the flux line at transverse position $\r_\perp$
within the
crystal is  $\psi_0^2(\r_\perp)$, where $\psi_0(\r_\perp)$ is the
normalized ground state eigenfunction of (2.7) [33].

When $B\gg B_{ c1}$, the potential is logarithmic, and we expand
$V_1(\r_\perp)$
about its minimum at $\r_\perp=0$ to find
$$
\left[-{T^2\over 2\tel}\;
\nabla_\perp ^2+{1\over 2}\;
kr_\perp^2\right]
\psi_0=\eps_0\psi_0
\eqno(2.8{\rm a})
$$
where (neglecting constants of order unity)
$$
\eqalignno{
k&\left.\approx {d^2V\over dr^2}\right|_{r=a_0}\cr
&\approx{\eps_0\over a_0^2}
&(2.8{\rm b})\cr}
$$
and $a_0$ is the mean vortex spacing. Equation (2.8a) is the Schroedinger
equation for a two-dimensional quantum oscillator, with $\hbar\rightarrow T$
and mass
$m\rightarrow\tel$. The ground state wave function is
$$
\psi_0(r_\perp)=
{1\over\sqrt{2\pi}\;r_*}\;
e^{-r^2/4r_*^2}
\eqno(2.9)
$$
with spatial extent
$$
r_*=\left({T^2a^2_0\over
\eps_0\tel}\right)^{1/4}\;.
\eqno(2.10)
$$
Melting occurs when $r_*=c_La_0$, where $c_L$ is the Lindemann constant, so the
melting temperature is
$$
T_m=c_L^2\sqrt{\eps_0\tel}\;a_0\;,
\qquad\qquad\qquad  (B_{ c1}\ll B\;\lot\;B_\times)
\eqno(2.11)
$$
in agreement with other estimates [27].
Note that $\eps_0$ and $\tel$ are themselves temperature dependent
so that Eq.~(2.11) is an implicit expression for $T_m(B)$, or, equivalently, a
melting field $B_m(T)$.

This result is only applicable when $B_m \ll
H_{c2}(T_m)=\phi_0/[2\pi\xi_\perp^2(T_m)]$ where
$\xi_{\perp}(T)$ is the temperature dependent in-plane coherence length
including the effects of critical thermal fluctuations
which make $H_{c2}(T)\sim(T_c-T)^{4/3}$ for $T$ sufficiently
close to the zero-field transition at $T_c$.
In this regime
$\lambda_\perp\sim(T_c-T)^{-1/3}$ and
the resulting ratio
$$
B_m/H_{c2}(T_m)={\rm const.}
$$
independent of magnetic field or anisotropy.
The logarithmic factor in $\tel$, Eq.~(2.4a), is thus of order unity
since $a_0/\xi_\perp(T_m)={\rm const}$. Such  critical effects were not
taken into account in calculations of the melting temperature prior to
Ref.~12.

Vortices in the crystalline phase will
travel a perpendicular distance $r_\perp^*$ in a ``time''  along the
$z$ axis  $\ell_z^0$, where [7]
$$
\eqalignno{
\ell_z^0&\approx r_*^2/(T/\tel)\cr
&\approx \sqrt{{\tel\over\eps_0}}\;
a_0\;.
&(2.12)\cr}
$$
A new high field regime arises when $\ell_z^0\;\lot\;d_0$, where $d_0$ is the
average spacing of the copper-oxide planes, i.e., for $B\;\got\; B_\times$,
with [12,24]
$$
B_\times\sim{\tel\over \eps_0}\;
{\phi_0\over d_0^2}\;,
\eqno(2.13)
$$
which is up to a logarithmic factor the same criteria discussed in the
previous section, $B_\times\sim(\gamma^2\phi_0/d_0^2)$.
Well above this field, the planes are approximately decoupled, and $T_m$ may be
estimated from
the theory of two-dimensional dislocation mediated melting
[34,12,24]
$$
T_m\approx 0.5 {\eps_0d_0\over
8\pi\sqrt 3}\;,
\qquad\qquad\qquad (B\gg B_\times)\;.
\eqno(2.14)
$$
where the prefactor 0.5 is a rough estimate of the effects of phonon
fluctuations [34]. Note that the convergence to Eq.~(2.14) for high
fields may be quite slow due to the very strongly divergent translational
correlation length at the melting temperature in two dimensions [12,24];
this causes $T_m(B)$ to appraoch Eq.~(2.44) only as $1/\ln^2(B/B_\times)$.

The estimate (2.11) also breaks down at low fields $B\;\lot\;B_{c1}$ where the
logarithmic interaction potential must be replaced by an exponential repulsion.
The two-dimensional harmonic oscillator
approximation can again be used with the replacement
$$
k\rightarrow {\eps_0\over\lambda_\perp^2}\;
e^{-a_0/\lambda_\perp}.
\eqno(2.15)
$$
The transverse wandering distance is now
$$
r_*\approx\left(
{T^2\lambda_\perp^2\over\eps_0\tel}\right)
^{1/4} e^{a_0/4\lambda_\perp}.
\eqno(2.16)
$$
and takes place over a longitudinal distance
$$
\ell_z^0=\sqrt{{\tel\over\eps_0}}\;
\lambda_\perp e^{a_0/2\lambda_\perp}.
\eqno(2.17)
$$
The low field melting temperature becomes
$$
T_m\approx c_L^2\sqrt{\eps_0\tel}
\;{a_0^2\over\lambda_\perp}
e^{-a_0/2\lambda_\perp}\qquad\qquad \qquad
(B\;\lot\;B_{ c1})\;,
\eqno(2.18)
$$
consistent with earlier predictions [7]. Although we have retained the
distinction
between $\tel$ and $\eps_0$ in these formulas, note from Eq.~(2.4b) that
$\tel\approx\eps_0$ in this regime.

The predictions (2.11), (2.14) and (2.18) are combined to give the reentrant
phase diagram for melting shown in Fig.~5 [35,12]. We note that when
fluctuations are relatively weak, so that melting occurs closer to $T_c$,
Eqs.~(2.11) and (2.18) must be solved self-consistently for $T_m$ using
the temperature dependence of $\epsilon_0$, $\tilde\epsilon_1$ and
$\lambda_\perp$. This procedure causes the maximum in the melting
temperature to bend downwards towards $T_c$ in materials like YBCO [12].
$B$ rather than $H$ is used for the
vertical axis since it is $B$ rather than $H$ which is fixed for the
thin, low aspect ratio
crystals usually studied in experiments. This figure is only schematic because
a first-order transition [2] at constant $B$ would lead to two-phase
coexistence with domain size limited by the strong interactions between
vortex tips as they exit the sample surface in thin
crystals.
Analytic estimates and boundaries for
melting in the various regimes are summarized in Table~1.
\medskip
\noindent
{\caps 2.2 Vacancy and Interstitial Unbinding Transitions}

Consider a defect such as the vacancy shown in Fig.~2. To estimate when its
free
energy changes sign, we need to determine the kink separation $\ell_z$ which
appears in Eq.~(1.3b). Up to factors of order unity, a similar estimate for
$T_d$ would apply to an interstitial vortex which, for example, hops
between the centers of the triangular cells of an Abrikosov crystal.
Interstitial
wandering would then take place on a honeycomb lattice, and the factor
$\ln 6$ in Eq.~(1.4b) would be replaced by $\ln 3$.

We suppose initially that $B_{ c1}\ll B\;\lot\;B_\times$,
so that the pair potential
is logarithmic with well-coupled CuO$_2$ planes. As shown in Sec.~3,
the defect line
energy is then proportional to the characteristic line
energy scale $\eps_0$ of the pair potential (2.2),
$$
\eps_d=c_d\eps_0\;.
\eqno(2.19)
$$
In Section 3, we determine the constant $c_d$
for various kinds of defects, and show that in fact
{\it interstitials}, rather than vacancies are favored energetically for
$B\gg B_{c1}$.

An interstitial surrounded by a triangular cage of nearest neighbors will
fluctuate in much the same way as the vortex in the hexagonal cage discussed in
the
previous section. To ``tunnel'' to one of its three neighboring triangular
plaquettes, it must overcome a barrier of order $\eps_0$. The energy
associated with one of these kinks or tunneling events is of order
$$
E_k\sim\sqrt{\eps_0\tel}\;a_0\;,
\eqno(2.20)
$$
from Eqs.~(2.12) and (2.1) since the ``kinks'' are spread over a distance
$\ell_z^0$.
This expression is just the WKB tunneling exponent which follows from the
``quantum mechanical'' analogy discussed above. The
line-density of kinks is then of order
$$
n_{{\rm kink}}\approx
{1\over\ell_z^0}\;
e^{-E_k/T},
\eqno(2.21)
$$
where $\ell_z^0$ should be the same order of magnitude as in Eq.~(2.12).
Similar estimates would apply to vacancies: Kinks arise in Fig.~2, for example
when
one of the flux lines surrounding an unoccupied site jumps over a barrier in
the lattice potential $U_\ell(\r)$ of order $\eps_0$ and essentially
changes places with the vacancy. The ``attempt frequency'' associated
with this event should again be of order $1/\ell_z^0$.

For both types of defects, we conclude that the spacing between kinks is of
order $\ell_z=n_{{\rm kink}}^{-1}$, or
$$
\ell_z\sim\ell_z^0
e^{E_k/T}\;.
\eqno(2.22)
$$
Upon substituting Eqs. (2.19) and (2.22) into (1.3b), we find a simple
self-consistent equation for $T_d$,
$$
T_d=c_1\sqrt{\eps_0\tel}\;a_0
\exp[c_2\sqrt{\eps_0\tel}a_0/T_d]
\eqno(2.23)
$$
where $c_1$ and $c_2$ are constants of order unity. It follows that
$$
T_d=c_3\sqrt{\eps_0\tel}\;a_0
\eqno(2.24)
$$
which has the same form as Eq.~(2.11) but with
$c_3$ a different numerical constant. Therefore
a supersolid phase
will occur whenever $c_3<c_L^2$. We attempt to estimate $c_3$ in Section~4.

When $B\gg B_\times$, defects can hop one lattice constant or more
between copper-oxide planes of spacing $d_0$,
and the defect path integral (1.3) must be
evaluated in a different limit. We now neglect the lattice potential
$U_\ell(\r_d)$ entirely and evaluate the functional integral which remains
by discretizing the path integral along $z$ in units of $d_0$. The result is
$$
F_d=\eps_dL-{T\over d_0}\ln
\left({2\pi B\over B_\times}\right)L
\eqno(2.25)
$$
which leads to a defect unbinding temperature
$$
T_d(B)\approx{\rm const.}\times
\eps_0d_0/\ln[2\pi B/B_\times]\;.
\eqno(2.26)
$$
Since $T_d\sim 1/\ln(2\pi B/B_\times)$
in this regime, while $T_m$ is  asymptotically
$B$-independent according to Eq.~(2.14), defects {\it must}
proliferate for $B\gg B_\times$. Equation~(2.26) agrees with the
unbinding transition of dislocation pairs from bound dislocation
quartets estimated in Ref.~13.

The result Eq.~(2.26) for the limit $B_\times\gg B$ can also be obtained
from a simple physical argument. In the limit of nearly decoupled
layers with a very weak Josephson coupling, the density of defects in each
layer is just
$n_d\sim e^{-E_d/T}$ where
$$
E_d=d_0\eps_d\;,
\eqno(2.27)
$$
is the energy of the defect in a single layer.
When the interactions between layers are weak, they can be estimated
perturbatively. If a vortex passing through two neighboring layers is
misaligned
by an amount $b$, the excess cost is of order $\gamma^2\eps_0b/d_0$
from the Josephson coupling between the layers. Naively, the cost of
misalignment of the vortex which makes up an interstitial ``line'' is
obtained by setting $b=a_0 n_d^{-1/2}$, the typical jog distance
of the defect between nearly decoupled layers.
This  is certainly an upper bound for the kink-free energy since fluctuations
of the other vortices in each layer will add a large incoherent part
to the phase difference
associated with the interstitial.
We conjecture that this will result in a reduction of the full
Josephson energy to
$$
E_I\sim{\gamma^2\eps_0a_0^2 n_d^{-k}\over d_0}
\eqno(2.28)
$$
with $k\le1$. The free energy per layer of the defects is then
$$
F_d\approx E_d n_d+T n_d\ln n_d+n_d E_I(n_d)\;.
\eqno(2.29)
$$
Minimizing with respect to $n_d$ yields the result that $F_d$
becomes negative, and hence $n_d$ positive, when $T>T_d$ with
$$
T_d\approx kE_d/\ln(B/B_\times)
\eqno(2.30)
$$
just the form of Eq.~(2.26).
Note that the {\it magnetic} interlayer coupling will only add an $n_d$
independent term to $E_I$ because of the screening due to relaxation
at the other vortices as discussed in Appendix~E.

The 2D defect energy $E_d$ is correctly obtained via Eq.~(2.27) from the
{\it 3D} calculations of $E_d$ described in this paper since in the absence
at Josephson coupling the vortices in 2D layers interact logarithmically
at all distances with the effective penetration length at long distances
only reduced by negligible $\CO(d_0/\lambda_\perp)$ corrections for small
$d_0$.

Finally, when $B\;\lot\;B_{ c1}$, the exponential interaction between vortices
leads to a different result. The lowest energy defect in this regime of
short-range interactions is presumably a vacancy, as is usual for solids with
short-range interactions. The characteristic defect line energy is now
$$
\eps_d={\rm const.}\times{a_0^2\over \lambda_\perp^2}\;\eps_0
e^{-a_0/\lambda_\perp}\;.
\eqno(2.31)
$$
The barrier to produce a kink in the trajectory of a vacancy or
interstitial has similar exponential dependence on $a_0$,
so the resulting
kink energy is
$$
E_k\sim\sqrt{\eps_0\tel}\;{a_0^2\over\lambda_\perp}\;
e^{-a_0/2\lambda_\perp}\;.
\eqno(2.32)
$$
Upon using Eq. (2.17) for $\ell_z^0$ in Eq. (2.22), we find via
Eq.~(1.30) that
$$
F_d=\eps_dL\left[
1-{\rm const.}\times {T\over E_k}\;
e^{-E_k/T}\right]
\eqno(2.33)
$$
which (provided the constant is not too small) changes sign for
$$
\eqalignno{
T_d&\approx {\rm const.}E_k\cr
&\approx{\rm const.}\times \sqrt{\eps_0\tel}\;
\left({a_0^2\over\lambda_\perp}\right)e^{-a_0/2\lambda_\perp}
&(2.34)\cr}
$$
in the limit of low fields. A supersolid phase is in principle possible
for low fields if $T_d<T_m$ as given by Eq.~(2.18). We conclude in Sec.~4
however, that $T_m<T_d$ for $B\;\lot\; B_\times$, which makes this
possibility rather unlikely.

Our estimates for interstitial or vacancy proliferation temperatures are
summarized
in Table~1. Combining these estimates for $T_d$ and $T_m$ (and
assuming that the supersolid is {\it un}favorable for small fields) leads to
the
schematic phase diagram shown in Fig.~5.
\vfill\eject
\noindent
{\caps 3. FORMATION ENERGY OF LINE DEFECTS}

In this section we discuss more precisely the defect line energy $\ep_d$ for
rigid
configurations of straight vortex lines. Numerical calculations are presented
for various types of defects in the limit $B_{ c1}\ll B\ll H_{c2}$.

\noindent
{\caps 3.1 Gibbs Free Energy of a Perfect Vortex Lattice}

Let $\vx (\ell)$ be the equilibrium position of the $\ell$th flux line. In
the London limit, $\kappa = \lambda_\perp/\xi_\perp\gg 1$, the pair potential
between two straight flux lines is given by Eq.~(2.2),
$$
V(r_{\ell \ell^\prime}) =
{\phi_0^2 \over 8 \pi^2 \lambda_\perp^2}
K_0 (r_{\ell \ell^\prime}/  \lambda_\perp)\;,
\eqno(3.1)
$$
where $r_{\ell \ell^\prime} = |\vx (\ell) - \vx (\ell^\prime)|$,
and $K_0$ is the modified Bessel function, with the asymptotic
behavior $K_0(x) \approx (\pi/2x)^{1/2} e^{-x}$ for large x and
$K_0(x) \approx - \ln (x/2) - \gamma$ for small $x$.

For $\H\;\Vert\;\B\;\Vert\;\c$  the Gibbs free energy per unit length $L$ and
unit
area $A$ is
$$
g = n \left( \epsilon_1 - {H \phi_0 \over 4 \pi} \right) +
n \ep_0
{\sum_\ell}^\prime K_0 (r_{0 \ell} / \lambda_\perp)\;.
\eqno(3.2)
$$
Here
$\epsilon_1 =
\left( {\phi_0 \over 4 \pi \lambda_\perp} \right)^2 \ln \kappa=\ep_0\ln\kappa$
is the energy per unit length of a single flux
line (ignoring an additive constant correction to $\ln\kappa$),
and $n=N/A$ is the number of lines per unit area.  The second
term in the Gibbs free energy represents the interaction
energies of the lines, while the first includes
the effect of the external magnetic field $H$ and favors
large values of $B$. $H$ plays the role of a pressure (or chemical
potential), which tends to increase the density of lines. The
prime denotes that self--interactions (the $\ell=0$ term) are omitted
from the summation.
The equilibrium magnetic flux density $B_{\rm eq} = n \phi_0$ is
obtained by minimizing the Gibbs free energy, i.e., by solving
$dG/dB = 0$. In the limit
$B\eq \gg \phi_0/\lambda_\perp^2$ one finds [36]
$$
H - H_{c1} = B_{\rm eq} + {\phi_0 \over 8 \pi \lambda_\perp^2}
\left[ \ln \left( { \phi_0 \over B_{\rm eq} \lambda_\perp^2 } \right) + C
\right] +
{\cal O} (B_{\rm eq}^{-1}) \;,
\eqno(3.3)
$$
where $H_{c1}={4\pi\ep_1\over\phi_0}$. The constant
$C = - \ln 4 \pi -2  + \gamma + A$ depends on the lattice
structure with  $\gamma \approx 0.5772..$ is Euler's constant.
For a hexagonal and square lattice we find
$A_6 \approx 0.0797107$, and $A_4 \approx 0.1008797$, respectively [36].
For a rectangular lattice with lattice constants $L_x$ and $L_y$ the value
of $A$ has to be calculated from
$$A = {\sum_{\vec R}}^\prime E_1 \left( {\pi \vec R^2 \over L_x L_y} \right)
      + {\sum_{\vec G}}^\prime { \exp \left[- \vec G^2 L_x L_y / 4 \pi \right]
                               \over \vec G^2 L_x L_y / 4 \pi}\;.
\eqno(3.4)
$$

The corresponding minimum value of the Gibbs free energy is
$$
G_{\rm eq} = - { B_{\rm eq}^2 \over 8 \pi} +
{ B_{\rm eq} \phi_0 \over 32 \pi^2 \lambda_\perp^2} +
{\cal O}(1) \;.
\eqno(3.5)
$$
We are interested in the formation energy of  ``point defects''
at constant line density $n$ corresponding to a fixed  external
magnetic field $H$. This later condition allows us to consider only the
interaction part of the free energy (3.2) for $N$ particles, i.e.,
$$
E_{N} = N \ep_0
{\sum_\ell}^\prime K_0 (r_{0\ell}/\lambda_\perp) \;,
\eqno(3.6)
$$
in determining defect energies. With Eqs. (3.2)--(3.5) one finds
$$
E_{N} = 2N \ep_0
\left\{
n \pi \lambda_\perp^2 + {1 \over 4}
\left[
\ln \left({1 \over n \lambda_\perp^2} \right) + C + 1
\right]
\right\}
\eqno(3.7)
$$

\noindent
{\caps 3.2 Definition of Defect Energies}

\noindent
{\it 3.2.1 Constant lattice spacing}

We follow Ref.~25 and
illustrate  the definition of defect energies for the case of a vacancy line.
Consider first a perfect lattice with $N+1$ flux lines confined to an area~$A$.
The corresponding
interaction energy is $E_{N+1}$. Removing a line at the origin without allowing
the flux lattice to relax reduces the interaction energy to
$$
E^\prime = {N+1 \over 2 } {\sum_\ell}^\prime V(r_{0\ell}) -
{\sum_\ell}^\prime V(r_{0\ell}) \;,
\eqno(3.8)
$$
i.e., we have subtracted the interaction energy of the removed center line
with the rest of the lines. Hence the (unrelaxed) defect energy for a vacancy
at constant lattice spacing is given by
$$
E_{\rm V}^{(2)} = E^\prime - E_{N+1} = - {2 \over N+1} E_{N+1} \;.
\eqno(3.9)
$$

\noindent
{\it 3.2.2 Constant line density}

The vacancies that occur in a superconductor
occur at constant line density or constant chemical potential, not
constant lattice constant.
Starting with a large perfect lattice containing $N$ flux lines in a fixed
area~$A$ one can imagine rearranging the flux
lines (with $N$ fixed)
in such a way that the resulting configuration is identical to the one
generated above by removing one line from a perfect lattice containing $N+1$
lines. Thus, the unrelaxed vacancy energy at constant line density is defined
by~[25]
$$
E_{\rm V}^u = E^\prime - E_N = E_{\rm V}^{(2)} +
\left( E_{N+1} - E_N \right) \;.
\eqno(3.10)
$$
Upon using  Eqs. (3.2)--(3.7) one gets for the Gibbs free energy
per unit length of a vacancy line in an {\it unrelaxed} lattice,
$$
G_{\rm V}^u = E_{\rm V}^u
- {\ep_0 \over 2}
\left[ \ln\left({1 \over n \lambda_\perp^2}\right) + C + 2 \right] \;.
\eqno(3.11)
$$
Note that we have neglected terms of the order $n/N=1/A$ since we
are interested in the thermodynamic limit with $n$ fixed and
$N \rightarrow \infty$.
The defect energy itself is $\ep_{\rm V}=G_{\rm V}\sim\ep_0
\left[\ln\left({\lambda_\perp\over a_0}\right)+{\rm const.}\right]$
which is the correct order of magnitude except for the logarithmic
divergence as $a_0/\lambda_\perp \rightarrow 0$. As we shall see, this
singularity
disappears when we allow the lattice to relax around the defect.

\noindent
{\caps 3.3  Elasticity Theory}

Before allowing static relaxations, we review the energies of the phonon
distortions which are involved.
When each vortex line is given
an arbitrary displacement $\u (\ell)$ from its equilibrium position
$\vx (\ell)$ the interaction (potential) energy of the vortex crystal is given
by
$$
E = {1 \over 2 A} {\sum_{\ell,\ell^\prime}}^\prime
V (|\vx(\ell\ell^\prime) + \u(\ell\ell^\prime) | ) \;,
\eqno(3.12)
$$
where $\vx(\ell\ell^\prime) = \vx (\ell) - \vx (\ell^\prime)$ and
$\u ({\ell\ell^\prime}) = \u (\ell) - \u (\ell^\prime)$ are the differences in
the lattice vectors and displacement vectors, respectively.
Expanding up to second order in the displacement fields yields
$$
E = E_N +
{1 \over 2 A} \sum_{\ell, \alpha}  \sum_{\ell^\prime, \beta}
\phi^{\alpha \beta} (\ell,\ell^\prime)
u^\alpha (\ell) u^\beta (\ell^\prime) + \quad ...  \;,
\eqno(3.13)
$$
where $E_N$ is the interaction energy of a rigid flux lattice. Here
$\alpha, \beta = x, y$ label the Cartesian components and
the dynamic matrix is given by
$$
\phi^{\alpha \beta} (\ell,\ell^\prime) =
\cases{- {\partial^2 V(r_{\ell \ell^\prime}) \over \partial x^\alpha
x^\beta} &for $\ell \neq \ell^\prime\;,$ \cr
\sum_{\ell \neq \ell^\prime}
{\partial^2 V(r_{\ell \ell^\prime}) \over \partial x^\alpha x^\beta}
&for $\ell = \ell^\prime\;.$ \cr}
\eqno(3.14)
$$
Upon defining the Fourier transform
$$
C^{\alpha \beta} (\vq)
= a_c \sum_{\ell} \phi^{\alpha \beta} (\ell,0)
e^{-i \vq \cdot \vx (\ell,0)} =
- \left( S^{\alpha \beta} (\vq) - S^{\alpha \beta} (0) \right) \;,
\eqno(3.15)
$$
with
$$
\eqalignno{
S^{\alpha \beta} (\vq)
 & =  a_c \sum_{\ell ( \neq \ell^\prime)}
{\partial^2 V(r_{\ell \ell^\prime}) \over \partial x^\alpha
\partial x^\beta} e^{-i \vq \cdot \vx (\ell,\ell^\prime)} \cr
&= a_c \lim_{\vx \rightarrow 0}
{\partial^2 \over \partial x^\alpha \partial x^\beta}
\left[ e^{- i \vq \cdot \vx} \sum_\ell e^{-i \vq \cdot ( \vx (\ell) - \vx )}
V \left( { \vx (\ell) - \vx } \right) -
V \left( { \vx } \right)
\right] \;, &(3.16)\cr  }
$$
the interaction energy can be written as
$$
E = E_N + {1 \over 2 A a_c^2}
\int_{\vq} u^\alpha (\vq)  C^{\alpha \beta} (- \vq) u^\beta (-\vq)
\;.
\eqno(3.17)
$$
Here $a_c = {\sqrt{3} \over 2} a_0^2=n^{-1}$ is the volume of the unit cell of
the
triangular lattice and we have introduced the short hand notation $\int_{\vq}
= \int {d^2 q \over (2 \pi)^2}$ for integrals over the in-plane wave
vector $\vq$.

In the limit of a very large London penetration depth, $\lambda_\perp \gg a_0$,
the dynamic matrix takes the form (see Appendix A)
$$
C^{\alpha \beta} (\vq) =
2\ep_0
\left\{ 2 \pi {q^\alpha q^\beta \over q^2} +
{a_c \over 8} \left[ \delta^{\alpha \beta}
q^2 - 2 q^\alpha q^\beta \right]
\right\}
\eqno(3.18)
$$
for small values of $q$.  Upon comparing with the usual continuum elastic
description of vortex solids [27], $C^{\alpha \beta} (\vec q)
= a_c^2 \left[ c_11 (\vec q) q^\alpha q^\beta +
               c_{66} (\vec q) \delta^{\alpha\beta} q^2
        \right]$,
we find that the bulk and shear moduli
are
$$
c_{11}(\vq)\approx
n^2{4\pi\ep_0\over q^2}-{n\ep_0\over 2}\;
\eqno(3.19)
$$
$$
c_{66}(\vq)
\approx {n\ep_0\over 4}\;.
\eqno(3.20)
$$
where $n=a_c^{-1}$ is the vortex density. In this limit of rigid, parallel
vortices we obtain no information on the $q_z$-dependence of these quantities
or on the tilt modulus $c_{44}$. Equation~(3.20) agrees with the shear modulus
first obtained by Fetter \etal \ [38]. Equation~(3.19) is correct for
$\lambda_\perp^{-1}<q_\perp<a_0^{-1}$, which encompasses a wide range of
length scales when $B\gg\phi_0/\lambda_\perp^2$.

\noindent
{\caps 3.4 Variational Calculation of the Relaxational Energy
of a Straight Vacancy Line}

In this section we calculate the relaxation energy of a straight vacancy
line by a variational approach. The presence of a vacancy causes a
distortion of the lattice described by the displacement field $\u (\ell)$.
Assuming the displacements $\u (\ell)$ to be small and  slowly
varying the relaxational energy for a vacancy can be obtained by
minimization with respect to $\u (\ell)$ of
$$
\eqalignno{
E_{\rm VR}^{(2)} ( \{ \u \} )
= & {1 \over 2 } \sum_{\ell, \ell^\prime} \phi^{\alpha \beta}
(\ell,\ell^\prime)
u^\alpha (\ell) u^\beta (\ell^\prime)\cr
&-  1  \sum_{\ell \neq 0} V_1^\alpha (\ell) u^\alpha (\ell) \cr
&-  {1 \over 2 } \sum_{\ell \neq 0} V_2^{\alpha \beta} (\ell)
u^\alpha (\ell)  u^\beta (\ell)\;,&(3.21)\cr}
$$
where
$$
\eqalignno{
V_1^\alpha (\ell) = & {\partial \over \partial x^\alpha } V (\vx) &(3.22)\cr
V_2^{\alpha \beta} (\ell) = &
{\partial^2 \over \partial x^\alpha \partial x^\beta} V(\vx) &(3.23)\cr}
$$
and the dynamical matrix $ \phi^{\alpha \beta} (\ell,\ell^\prime)$ is
defined in Eq.~(3.14).

The first term in $ E_{\rm VR} ( \{ \vu \} ) $ is the elastic
energy of the lattice distortion caused by the vacancy.
It is overcounts the energy of the missing center line with
the rest of the crystal. This contribution is subtracted by the
second and third term in Eq.~(3.21), which is just a Taylor
expansion of $ \sum_{\ell \neq 0} V (\mid \vx (\ell) + \vu (\ell) - \vx (0)|)$,
where $\vx (0) = 0$ is the position of the vacancy in the lattice.

In Fourier space, we have
$$
\eqalignno{
u^\alpha (\vq) = &
a_c \sum_\ell u^\alpha (\ell) e^{-i \vq \cdot \vx (\ell)}
&(3.24)\cr
C^{\alpha \beta} (\vq) = &
a_c \sum_\ell \phi_{\alpha \beta} (\ell,0) e^{-i \vq \cdot \vx (\ell)}
&(3.25)\cr
V_1^\alpha (\vq) = &
a_c \sum_{\ell \neq 0} V_1^\alpha (\ell) e^{-i \vq \cdot \vx (\ell)}
&(3.26)\cr
V_2^{\alpha \beta} (\vq) = &
a_c \sum_{\ell \neq 0} V_2^{\alpha \beta} (\ell) e^{-i \vq \cdot \vx (\ell)}
&(3.27)\cr}
$$
and the vacancy relaxation energy becomes
$$
\eqalignno{
E_{\rm VR}^{(2)} ( \{ \vu (\ell) \} )
= & {1 \over 2  a_c^2} \int_q u^\alpha (\vq) u^\beta (- \vq)
C^{\alpha \beta} (-\vq) \cr
- & {1 \over  a_c} \int_q  V_1^\alpha (\vq) u^\alpha (-\vq) \cr
- & {1 \over 2  a_c} \int_q \int_k V_2^{\alpha \beta} ( \vq - \vk )
u^\alpha ( \vk ) u^\beta (- \vq ) \;. & (3.28)\cr}
$$
Following Ref.~25, we make a variational Ansatz for the lattice
distortion field,
$$
u^\alpha (\vq) = i a_c { q^\alpha \over q^2} f ( q )\;,
\eqno(3.29)
$$
where $f(q) = 1 + c q + d q^2 $ with $c$ and $d$ as variational parameters.
As discussed in Appendix B, the constraint $f(0)=1$ is enforced by the long
range
potential. $V_1^\alpha (\vq )$ and $V_2^{\alpha \beta} (\vq )$ can be
expressed in terms of the dynamic matrix $C^{\alpha \beta} ( \vq )$
$$
V_1^\alpha (\vq) =  a_c \sum_{\ell \neq 0} V_1^\alpha (\ell)
e^{-i \vq \cdot \vx (\ell)} =
- - i {\partial \over \partial q_\gamma} V_2^{\alpha \gamma} (\vq) \;,
\eqno(3.30)
$$
$$
V_2^{\alpha \beta} (\vq) = a_c \sum_{\ell \neq 0} V_2^{\alpha \beta} (\ell)
e^{-i \vq \cdot \vx (\ell)} = - C^{\alpha \beta} (\vq) +
V_2^{\alpha \beta} (\vq = 0) \;,
\eqno(3.31)
$$
where we have used the asymptotic form of the potential, $\lim_{\lambda_\perp
\to \infty} V(x) = - \ln x$, to derive equation~(3.30).  This gives
$$
V_1^\alpha (\vq) =
2i \eps_0\, \left( {2 \pi \over q^2} - {a_c \over 2} \right) q^\alpha \;,
\eqno(3.32)
$$
and for the above Ansatz for the displacement field one can take
$$
V_2^{\alpha \beta} (\vq)  = - C^{\alpha \beta} (\vq ) \;.
\eqno(3.33)
$$
Upon using the long wavelength approximation (2.18) for
$C^{\alpha\beta}(\vq)$, one finds
$$
E_{\rm VR}^{(2)} ( \{ \vu (\ell) \} )  = 2\eps_0
\left( I_1 + I_2 + I_3 \right)
\eqno(3.34)
$$
with $I_i$ given in Appendix B.
The variational calculation for $c$ and $d$ yields (see Appendix~B)
$$
c =  {1 \over 7 k_d}      \approx   0.161 \sqrt{a_c} \;,
\eqno(3.35)
$$
$$
d = -{10  \over 7 k_d^2}  \approx - 0.114  a_c       \;,
\eqno(3.36)
$$
where $k_d=\sqrt{4\pi/a_c}$ is a cutoff which preserves the area
of the underlying hexagonal Brillouin zone.
Hence, the elastic relaxation contribution to the vacancy energy
density at constant lattice spacing is
$$
E_{\rm VR}^{(2)} = {\eps_0\over 2}\,
\left[\ln\left(1\over k_d^2\lambda^2_\perp\right)+{265\over252}\right]
\eqno(3.37)
$$

The {\it total} vacancy energy density at constant density instead of
constant lattice constant, is, following the considerations of Sec.~3.2,
$$
G_V=E^{(2)}_{\rm VR}+E_V^{(2)}+E_{N+1}-E_N\;.
\eqno(3.38)
$$
The last three terms are given by Eq.~(3.11), which, when combined with
(3.37), leads to
$$
G_V={\eps_0\over2}\,
\left[-\gamma-A_6+{265\over252}\right]\;.
\eqno(3.39)
$$
Our variational vacancy line-energy is thus $\epsilon_V=G_V$, or
$$
\epsilon_V=0.1973\eps_0\;.
\eqno(3.40)
$$
Note that the logarithmic divergences as $a_0/\lambda_\perp \to0$
in Eqs.~(3.11) and (3.37) have cancelled to yield a finite result.

A heuristic ``back of the envelope'' argument for the vacancy energy can also
be constructed: assume that the phonon displacement in real space for a
vacancy at the origin takes the isotropic form
$$
\vu(\ell)=-{\Omega_0\over2\pi}\,{\vx(\ell)\over x^2(\ell)}\;,
\eqno(3.41)
$$
consistent with a six-fold symmetry of this defect site [37]. The parameter
$\Omega_0$ is the area change induced in the flux line lattice.
To keep the {\it density} of flux lines fixed (as is appropriate for
$B\gg\phi_0/\lambda^2_\perp$), we take $\Omega_0=a_c$, the area of one unit
cell to
cancel the vacancy energy.
Since $\vnab\cdot\vu=0$ and $\partial_z\vu=0$, the only contribution from
the continuum elastic free energy per unit length comes from the wave-vector
independent shear modulus term,
$$
\epsilon_V=\int d^2x c_{66} u^2_{ij}\;,
\eqno(3.42)
$$
where $u_{ij}={1\over 2}\,(\partial_iu_j+\partial_ju_i)$, $i=1,2$ and [38]
$c_{66}={1\over 4}\,n_0\eps_0$. In  Fourier space, this
expression becomes
$$
\epsilon_V=c_{66}\int{d^2q\over(2\pi)^2}\,|u_{ij}(\vq)|^2\;,
\eqno(3.43)
$$
with $u_{ij}(\vq)=-a_cq_iq_j/q^2$. Upon using Eq.~(3.20) and imposing
a circular Brillouin zone of radius $k_d=\sqrt{4\pi/a_c}$, we find
finally
$$
\epsilon_V=\eps_0/4\;,
\eqno(3.44)
$$
an estimate only
25\% greater than Eq.~(3.40), and very close to the numerical
value for the relaxed vacancy configuration with six-fold symmetry [37].

\noindent
{\caps 3.5 Numerical Calculation of the Defect Energies}

Interstitial defects generally occupy lattice sites of low symmetry,
so that analytic methods become quite laborious. In this section we
describe numerical calculations of the various defect energies which
do not require the approximations used above. Our goal is to calculate
the defect energies for an infinite system.

In the computer simulations,
we have not an infinite system but a system with a large but finite
block of particles with periodic boundary conditions. In order to handle
the  long--range logarithmic interaction between the flux lines we use the
the Ewald sum technique, which amounts to including the interaction
of the flux line with all its periodic images [39].
For comparison with the computer results, one must therefore calculate
not the energy of a single defect in an infinite system, but the energy per
block of an periodic array of defects in an infinite system.
After correcting for these ``image'' defects one can extract the desired
energy  of a single defect. In practice we choose the system large enough
that the  interaction of the defects with its periodic images can either be
neglected or calculated by means of linear elastic theory.

The interaction energy of the flux line lattice per unit length in the
$z$ direction is
$$
E = {E_0 \over 2}
       \sum_{\ell,\ell^\prime} K_0 (r_{\ell \ell^\prime} / \lambda_\perp) \;,
\eqno(3.45)
$$
where our energy unit is $E_0 = 2 \eps_0$.
This sum over the infinite lattice can also be written as
$$
E  = {E_0 \over 2}
       \sum_{\ell,\ell^\prime \, \epsilon \, {\rm box}}
       \phi (r_{\ell \ell^\prime}/\lambda_\perp) \;,
\eqno(3.46)
$$
where the summation is over all particles within the basic box
of size $(L_x,L_y)$. The effective pair potential within the
basic box
$$
\phi (r_{\ell \ell^\prime}/\lambda_\perp) =
\mathop{{\sum}'}_\vR K_0 (\mid \vx(\ell \ell^\prime) + \vR \mid /\lambda_\perp)
\eqno(3.47)
$$
represents the interaction energy between a flux line at position
$\vx (\ell)$ and one at position $\vx (\ell^\prime)$ together with all its
images at positons $\vx (\ell^\prime) + \vR$.
The sum over $\vec R$ runs over all simple cubic lattice points,
$\vec R = (\ell L_x, m L_y)$ with $\ell$, $m$ integers. This vector
reflects the shape of the basic box. The prime on the summation sign
of Eq.~(3.47) indicates that we omit
the term $\vR = 0$ for $\vx (\ell\ell^\prime) = 0$.

For a numerical calculation equation (3.47) is not a suitable
starting point because of its poor convergence properties. Therefore,
we  make use of Ewald's summation method [39].
Using the integral representation of the modified Bessel function $K_0$
and Ewald's generalized theta function transform (see Appendix~A)
we find for $\vx \neq 0$ in the limit of large $\lambda$
$$
\eqalignno{
\phi(\vx/\lambda)& =
{1 \over 2} \sum_\vR E_1 \left( { (\vx-\vR)^2 \pi \over L_x L_y \delta}
\right)\cr
&\qquad + {1 \over 2} {\sum_{\vG}}^\prime e^{-i \vG \cdot \vx} \,
 {\exp \left[ -L_x L_y G^2 \delta /4 \pi \right] \over
  L_x L_y G^2 / 4 \pi} \cr
&\qquad +{2 \pi \lambda^2 \over L_x L_y} - {\delta\over 2} \;,&(3.48)\cr}
$$
where $E_1 (x) = \int_0^\infty dy y^{-1} e^{-y}$ is the exponential integral
function. The vectors $\vG = {2 \pi} (m/L_x,n/L_y)$ with $m$ and $n$ integers
index
the square lattice reciprocal to the lattice of image lines.
Note that this result is valid for any choice of the Ewald separation
parameter $\delta$. It can be used as a numerical check and to optimize
the convergence properties of Eq.~(3.48). We choose $\delta = 1$.

For $\phi(0)$ we get
$$
\eqalignno{
\phi(0)& =
{1 \over 2} {\sum_{\vR}}^\prime
E_1 \left(\pi R^2 \over L_x L_y\right) +
{1 \over 2} {\sum_{\vG}}^\prime
{\exp \left[-G^2 L_x L_y / 4 \pi \right] \over G^2 L_x L_y / 4 \pi}
+{2 \pi \lambda^2_\perp \over L_x L_y} - {1 \over 2}
+{1 \over 2} \ln {L_x L_y \over 4 \pi \lambda^2_\perp} + {\gamma \over 2}\cr
& = {2 \pi \lambda^2_\perp \over L_x L_y} +
{1 \over 2} \ln \left( {L_x L_y \over \lambda_\perp^2} \right) +
{1 \over 2} \left( \gamma - 1 - \ln 4 \pi + A_{\rm rect} \right) \;.
&(3.49)\cr}
$$
The value of the Ewald sum $A_{\rm rect}$ depends on the shape of the
basic box (compare with Eq.~(3.4)).
We choose the rectangular basic box such that $L_x = 5 a_0 \cdot t$
and $L_y = 3 \sqrt{3} a_0 \cdot t$ with $t$ integer. Then one
gets from a numerical evaluation of the Ewald sums
$A_{\rm rect} \approx 0.1018412$ and the
interaction energy in units of $E_0$ becomes
$$
E = {N^2 \over 2} \, { 2 \pi \lambda_\perp^2 \over L_x L_y} +
{N \over 4} \left[ \gamma - 1 - \ln 4 \pi + A_{\rm rect} +
\ln \left(L_x L_y\over\lambda_\perp^2\right)\right]
+{1 \over 2} {\sum_{\ell,\ell^\prime}}^\prime {
\hat \Phi} (r_{\ell \ell^\prime} / \lambda_\perp)
\eqno(3.50)
$$
with the effective pair potential
$$
\hat\Phi (r_{\ell \ell^\prime}) =
{1 \over 2} \sum_{\vR} E_1
\left((\vx - \vR)^2 \pi \over  L_x L_y \right) +
{1 \over 2} {\sum_{\vG}}^\prime e^{-i \vG \cdot \vx} \,
{\exp \left[ -L_x L_y G^2  / 4 \pi \right] \over
L_x L_y G^2 / 4 \pi}  - {1 \over 2} \;.
\eqno(3.51)
$$
Upon comparing Eq.~(3.50) with the interaction energy of a perfect flux
line lattice, Eq.~(3.7), one gets for a rectangular basic box
$$
{1 \over 2} {\sum_{\ell,\ell^\prime}}^\prime {\hat \Phi} (r_{\ell \ell^\prime})
 =  {N \over 4} \left( - \ln N + A_{6} - A_{\rm rect} \right)\;,
 \eqno(3.52)
$$
where $N=30\cdot t^2$ is the number of flux lines in the box of
size $(L_x,L_y) = (5, 3 \sqrt{3}) t a_0$.

For the numerical calculations it is sufficient to consider only those parts
of the interaction energy, which explicitely depend on the coordinates
of the flux lines. Hence we consider the quantity
$$
\hat E= {1\over2} \sum_{\ell,\ell^\prime} \hat\Phi (r_{\ell \ell^\prime}) \;.
\eqno(3.53)
$$
The force between flux lines at a distance $\vx(\ell \ell^\prime)$ is given by
$$
\vF (\vx_{\ell \ell^\prime}) = \vF_{\ell \ell^\prime} =
-{\partial \over \partial \vx(\ell \ell^\prime)}
\hat\Phi (r_{\ell \ell^\prime}) \;.
\eqno(3.54)
$$
Using the above expression for the interaction potential ${\hat \Phi}$
one can write
$$
\eqalignno{
\vF (\vx) = {1 \over 2} \Biggl\{ &\sum_\vR {2 \pi \over L^2} \,
{\exp \left[ -\mid \vx  - \vR \mid^2 \pi / L_x L_y \right] \over
\mid \vx  - \vR \mid^2 \pi / L_x L_y } \, \left( \vx  - \vR \right) \cr
+&{\sum_\vG}^\prime \vG \sin \left( \vG \cdot \vx  \right) \,
{\exp\left[ G^2 L_x L_y / 4 \pi \right]\over G^2 L_x L_y / 4 \pi}\Biggr\} \;.
&(3.55)\cr}
$$
These forces are most efficiently calculated together with the energy (in
one subroutine).

We now explain the algorithm by which the formation energies of the
various types of defect lines are determined.
We start with an initial configuration $\{ \vx^{(0)} (i) \}$ which,
after relaxation, will contain the desired defect.
This is easily achieved by adding or removing lines from
the perfect hexagonal lattice. This procedure leads to the defect energies at
constant lattice constant, and hence must be corrected as in Eq.~(3.38) to
produce energies at constant vortex density. The corresponding initial
configuration for a vacancy and a variety of other defects are shown in
Fig.~6.

The lattice relaxation process was performed by standard methods adapted
from molecular dynamics simulations [40]. For advancing the positions
and velocities of the flux lines we implemented an artificial dynamics via
a leapfrog algorithm
$$
\vv_\ell(t+{1 \over2} \delta t) = \vv_\ell(t-{1 \over 2}\delta t) +
                                  \delta t \, \va_\ell(t)
\eqno(3.56)
$$
$$
\vx_\ell(t+\delta t) = \vx_\ell(t) + \delta t \,
                                     \vv_\ell (t + {1\over 2}\delta t)
\eqno(3.57)
$$
The acceleration $\va_\ell = \vf_\ell / m$ of the $\ell$th flux line is
calculated from the forces $\vf_\ell = \sum_{\ell^\prime \neq \ell}
\vF_{\ell\ell^\prime}$. In each step of the
iteration we calculate the kinetic and potential energy, and the forces
using Eqs.~(3.50), (3.51) and (3.55).  This procedure is repeated
until an equilibrium configuration is reached. The ``mass'' $m$ and the
time step $\delta t$ were chosen to accelerate the convergence to equilibrium.
Equilibrium here means that the forces on the flux lines become zero. Hence
the method is capable of finding not only minima but also some saddle point
configurations, at least for initial states with high symmetry.

In order to test the accuracy of our method we calculated the energy of
a perfect flux line lattice with $N= 30 \times t^2$ particles for
$t = 1, \,2, \, ... ,\, 7$, and compared them with the exact result
Eq.~(3.52). Our results are summarized in Table~2. The relative
difference is less than $10^{-5}$ and can mainly be attributed to the
inaccuracy in the numerical approximation we have used for the exponential
integral function [41].

Starting from the initial configurations shown in Fig.~6 we have
determined the relaxed configurations using the algorithm described
by Eqs.~(3.56) and (3.57) for various system sizes. From this analysis
we can extrapolate to the defect formation energies of an infinite system.

For the vacancy we find that starting from the initial configuration
in Fig.~6, which has the six--fold symmetry of the hexagonal lattice,
the lattice first relaxes into a saddle point configuration, which
possesses the full symmetry of the lattice. This configuration, however,
is unstable
with respect to a compression along one of the three axis connecting
the nearest neighbors at the vacancy. It finally relaxes into a configuration
of lower symmetry (see Fig.~7). Note, that there are three equivalent
orientations of this relaxed vacancy configuration. In Table~3 we
summarize the formation energies of the stable and saddle point configuration
of the vacancy for different system sizes.

The edge interstitial relaxes starting from the initial configuration in
Fig.~6 into a saddle point configuration shown in Fig.~8. From the numerical
simulation we find  that this configuration is unstable with respect to
small amplitude ``buckling'' perpendicular to the edge. The edge interstitial
relaxes into a ``buckled configuration'', which is identical to the relaxed
centered interstitial configuration, also shown in Fig.~8.
These results are analogous to findings by Cockayne and Elser for the
two--dimensional Wigner crystal [25], where the edge interstitial is
also unstable with respect to ``buckling'' perpendicular to the edge
of the triangle. However, the time required for this
relaxation process is much larger than the relaxation time from an initial
edge interstitial configuration constrained by symmetry to go to a
final edge state. Thus, the  interstitial appears to occupy a
very flat minimum in configuration space.

In Table~3 we have listed the defect energies for a vacancy and several types
of interstitials  for various system sizes. Whereas there is a clear energy gap
between vacancies and interstitials, the interstitial energies are all very
close. The energy differences are less than 1\%! The system size dependence
for the centered edge, and split centered interstitials and two
types of vacancies are shown in Figs.~9a
and 9b, respectively.

The lattice conformations resulting from relaxing an edge interstitial and
a split edge interstitial initial configuration (constrained now
{\it not} to buckle as in Fig.~8) are shown in Fig.~10
(also shown by filled circles is the perfect hexagonal lattice). As can
be inferred from this figure the configurations essentially differ only by
a parallel shift along the edge of the triangle. Because the energies
are quite close (see Table 3) we conclude that gliding of this type of
defect along the direction defined by the edge of the triangle (in the
absence of buckling) must be
a low energy excitation. The barrier for this motion is presumably of the
order of the difference in energy of those configurations, i.e.
$\Delta_{\rm glide} \approx 10^{-5}E_0$!
After very long relaxation times both types of initial conformation, i.e.,
edge and split edge interstitial, may in fact finally relax into the same
final configuration.

We conclude that interstitials, rather than vacancies, are clearly
favored at high fields and that the centered interstitial is the most likely
candidate for producing a supersolid in this regime.
Since the differences between the energies of the various interstitials  are so
small, an interstitial will have substantial extra entropy, lowering its
free energy even further.

\noindent
{\caps 3.6 Interaction between Point Defects}

Following Cockayne and Elser [25], we can use defect energies for different
system
sizes
to infer the distance  dependence of the interaction energies.
As explained at the beginning of section 3.4, periodic boundary conditions
always yield a rectangular superlattice of defects. Since we are changing
only the size of the big box $(L_x, L_y)$ and not its shape, the system size
dependence of the formation energy should scale the same way as the radial
dependence of the interaction between two single defects for large $L_x$ and
$L_y$.

{}From the system size dependence of the centered and edge interstitial
energies, plotted
in Fig.~9a, one can infer that both defects show an attractive interaction at
distances larger than 5 lattice spacings. Over the range
studied $5\le r\le 35$ the interaction approximately
scales as $r^{-1}$ for centered interstitials and as $r^{-2}$ for edge
interstitials. From a nonlinear least square fit over the limited range  we get
$E_{\rm int}^{\rm EI} \approx 0.0742 - 0.0787*N^{-1.09}$
corresponding to $r^{-2.18}$ and
$E_{\rm int}^{\rm CI} \approx 0.0732 - 0.0078*N^{-0.61}$
corresponding to $r^{-1.22}$ for the edge and centered
interstitial,respectively.
As we shall see, these are definitely {\it not} the correct asymptotic long
distance behaviors.

We have also analyzed numerically the distance  dependence of the interaction
between two
centered interstitials at smaller distances.
In order to calculate this energy we have started with an initial
configuration, that contains two centered interstitials at a distance $d$
(measured in units of the lattice constant $a$)
in a perfect hexagonal lattice containing $N = 480$ flux lines. The
interaction energy is found by subtracting from the resulting relaxation
energy the energy of two isolated single centered interstitials. The results
for two different directions are displayed in Fig.~11a. If the vector
connecting the two centered interstitials points along one of the unit vectors
of the primitive cell of the hexagonal lattice, the interaction is attractive
up to $d  \approx 3\,a$ and becomes repulsive for larger distances.
In the direction perpendicular to
one of the unit cell vectors we find a minimum in the interaction potential
at a distance of approximately two lattice vectors. (Note that distances in
Figs.~11 are the distances between the defects in the initial configuration).
The system size difference discussed earlier indicates that, roughly, some
angular
average of the  interaction is attractive for $r>5a$.

In Fig.~9b we have plotted the system size dependence of the symmetric and
``crushed'' vacancy configurations. Whereas for a  symmetric vacancy
configuration the formation energy increases with increasing number of
flux lines $N$, it decreases for the ``crushed'' vacancy configuration.
{}From a nonlinear least square fit to the data in Fig.~9b we find
$E_{\rm int}^{\rm V2} \approx 0.1076 - 0.312*N^{-1.15}$
corresponding to $r^{-2.3}$ and
$E_{\rm int}^{\rm V6} \approx 0.125 + 0.0576*N^{-0.64}$
corresponding to $r^{-1.28}$ over the limited range $5\le r\le 35$ for   the
symmetric and ``crushed'' vacancy,
respectively. Hence we conclude that symmetric vacancies appear to have an
attractive
interaction at relatively long distances while  ``crushed'' vacancies,
which have the lower formation
energy
repel each other at comparable distances.

For smaller distances we have performed calculations analogous to those for the
centered
interstitial. The results for the stable ``crushed'' vacancy (V2) for
the interaction along (solid line) and perpendicular (dashed line) to an edge
of the unit cell are shown in Fig.~11.b.
Whereas the interaction energy is attractive for all distances less than $11 a$
along the directions perpendicular to the edges of the unit cell, it is
attractive for small distances and becomes repulsive for distances larger than
$d \approx 5 a$ for the interaction along the edge of the unit cell. This has
to be compared with the roughly  angular averaged
attractive interaction for $d > 5 a$ obtained
from the finite size analysis.

It is not possible to study the interaction between symmetric vacancies at
short
distances. This is because the anisotropy of the stresses induced by the
interactions causes the vacancies to deform to the anisotropic crushed
vacancy configuration which has lower energy. The symmetry axis of the
resulting crushed vacancies is parallel to their separtion vector.

It is instructive to compare these results with those obtained from linear
elasticity theory which should be valid for very large separations. It can be
shown that the $\ln r$ interactions between unrelaxed defects will be
completely screened by the relaxation of the other vortex lines. This is
already
evident in the calculations in section~3.4 for a single vacancy: the vortex
lines far from the vacancy relax just so as to cancel the overall ``charge''
of the vacancy. The long distance interactions between defects thus have the
same
form as for short-range interactions. These depend on the symmetries of the
defects involved.

For defects with three- or six-fold symmetry---the symmetric vacancies and
centered interstitials---the interactions are exponentially small in the
absence of anisotropy of the elastic interactions with the anisotropy
associated with the six-fold symmetry [which appear at order $q^4$ in
the elastic matrix $C^{\alpha\beta}(\vec q)$ of Eq. (3.17)], the interactions
will decay as $\cos 6\theta/r^4$ with $\theta$ the angle between a lattice
vector and the inter-defect separation vector. The sum over all the
inter-defect
interactions with the periodic boundary conditions used in the numerical
computations will thus almost cancel as the systems used are almost square with
$L_y^2/L_x^2=25/27$. The resulting asymptotic $L$ dependence of the
vacancy energy should thus have a leading $1/L^4$ term with a
small coefficient whose sign depends on details of the $q$ dependence of the
elastic matrix and the stresses induced by the vacancy which we have not
calculated.

The interactions between defects with only a two-fold symmetry axis-edge
and split-centered interstitials and crushed vacancies are longer range
since they would decay slowly even in an isotropic elastic medium. The
interactions will have two comparable contributions at long distances.
$$
E_{\rm int}\approx
{K_2\cos 2\phi\over r^2}+
{K_4\cos 4\phi\over r^2}
\eqno(3.58)
$$
where $\phi$ is the angle between the {\it symmetry axis} of the defect and
the separation vector. The coefficient $K_4$ is always positive (i.e.,
repulsive) while the sign of $K_2$ depends on the stresses induced by
the defects. In the almost square system with periodic boundary
conditions used in the numerical calculations, the $\cos 2\phi$ term will
almost cancel and the leading size dependence of the defect energies will be
$1/L^2$ with a sign $\cos4\alpha$ which depends on the cosine of the
angle $\alpha$ between the $x$ direction and the symmetry axis of the
defects.  At long distances, however, the actual sign of the interaction
between two isolated defects will depend on the direction of the
defect separation in a different way if $|K_2|>K_4$.

The apparent distance dependence seen in the numerical calculations is a fit
over
limited range of distance and presumably indicates that the asymptotic
behavior discussed above has not been reached.

\vfill\eject

\noindent {\caps 4. EQUILIBRIUM DESCRIPTION OF THE SUPERSOLID PHASE}

We now assume that centered interstitials are energetically preferred (as found
for zero temprature in
Sec.~3), and study the low lying excitations via a simple tight-binding
model. We find a more quantitative estimate of the transition temperature
$T_d$,  discuss the nature of the transitions at $T_d$ and $T_m$
and compare the properties of the solid, supersolid and liquid phases.

\noindent
{\caps 4.1 Tight-binding Model and the Transition Temperature}

As discussed in Sec. 2, the defect proliferation temperature is determined
by the large $L$ behavior of the partition function~(1.2). Equation~(1.2)
may also be written as an integral over a quantum mechanical matrix element
analogous to Eq.~(2.6),
$$
\CZ_d=\int d^2\vr_i \int d^2\vr_j
\left\langle\vr_f| e^{-\CH_d L/T}|\vr_i\right\rangle,
\eqno(4.1)
$$
where the integral is over entry and exit points $\vr_i$ and $\vr_j$ for the
interstitial and $\CH_d$ is an effective
quantum Hamiltonian describing the
transfer matrix. We  adopt a low-temperature perspective and use for
$\CH_d$ a tight-binding model which assumes that the centered interstitial sits
at the sites of a simple honeycomb lattice. As shown in Fig.~13, the
honeycomb lattice may be viewed as two interpenetrating triangular lattices
with sites $\{\vr_A\}$ and $\{\vr_B\}$, connected by displacement
vectors $\{\vec \delta_i,\ i=1,2,3\}$ with
$$
\vec \delta_1=b(0,1)\;,\qquad
\vec \delta_2=b\left({\sqrt3\over 2},\;
-{1\over 2}\right)\;,
\qquad
\vec \delta_3=b\left(-{\sqrt3\over 2},\;-{1\over 2}\right)
\eqno(4.2)
$$
and $b={\sqrt3\over 3}\;a$ with $a_0$ the triangular lattice
constant.

The interstitial can ``tunnel'' from one site to
another by passing through an edge (interstitial)
state, which according to Table~3, has an
energy nearly degenerate with the centered interstitial. The tight-binding
Hamiltonian is then
\vfill\eject
$$
\eqalignno{
\CH_d&=u_0
\left[\sum_{\vr_A}
|\vr_A\ra\;\la\vr_A|+
\sum_{\vr_B}|\vr_B\rangle\;
\langle\vr_B|\right]\cr
&\qquad -t\left[
\sum_{\vr_A,\vec \delta_A}
|\vr_A\ra\;\la\vr_A+\vec \delta_A|+
\sum_{\vr_B,\vec \delta_B}
|\vr_B\ra\;\la r_B+\vec \delta_B|\right]
&(4.3)\cr}
$$
where $|\vr_A\ra$ and $|\vr_B\ra$ are normalized states on the two sublattices,
$\{\vec \delta_A\} =\{\vec \delta_i\}$ and $\{\vec \delta_B\}=\{-\vec
\delta_i\}$. Here
$\u_0$ is a site energy, and $t$ is a tunnelling matrix element.

The eigenvectors of (4.3) are the linear combinations $\psi_\pm(\vk)\equiv
{1\over\sqrt2}(|\vk,A\ra\pm|\vk,B\ra)$ of normalized plane wave states,
$$
|\vk,A\ra={1\over\sqrt{N_0}}
\sum_{\vr_A}\;
e^{i\vk\cdot\vr_A}
|\vr_A\ra,\qquad
|\vk,B\ra={1\over\sqrt{N_0}}
\sum_{\vr_B}\;
e^{i\vk\cdot\vr_B}
|\vr_B\ra,
\eqno(4.4)
$$
where $N_0$ is the number sites in one sublattice and $\vk$ is confined to
a hexagonal Brillouin zone. The eigenvalues consist of two bands,
$$
\ep_\pm(\vk)=u_0\mp|t|\sqrt{
3+2[\cos(\vk\cdot\vec \delta_1)+
\cos(\vk\cdot\vec \delta_2)+
\cos(\vk\cdot\vec \delta_3)]}
\eqno(4.5)
$$
which are nondegenerate except at the zone corners. The lowest eigenvalue
occurs
in $\eps_+(k)$ at $\vk=0$,
$$
E_0(T)=u_0-3|t|\;,
\eqno(4.6)
$$
and dominates the partition sum (4.1) in the limit $L\rightarrow\infty$. The
defect unbinding temperature $T_d$ is determined by the condition $E_0(T_d)=0$.
To proceed further, we need to determine $u_0$ and $t$.

We assume the tunnelling process between sites of the honeycomb lattice
can be modeled by a one-dimensional quartic potential in the coordinate
$x$ connecting two honeycomb lattice sites,
$$
V(x)={
\tilde\ep_d\omega^2\over 2b^2}\;
\left( x-{1\over 2}\;b\right)^2
\left( x+{1\over 2}\; b\right)^2
\eqno(4.7)
$$
where the potential vanishes at the lattice sites. The frequency $\omega$ is
fixed by equating the maximum to $\Delta\ep>0$, the energy difference
between the edge and centered interstitials in Table~2,
$$
\omega=4\sqrt2\sqrt{
{\Delta\ep\over\tilde\ep_d}}\;
{1\over b}\quad.
\eqno(4.8)
$$
A standard quantum mechanical calculation [43] then gives
$$
E_\pm=T\omega\left[
{1\over 2}\;\pm\; 2
\sqrt{{3\over 2\pi}}
\left({S_0\over T}\right)^{1/2}
e^{-S_0/T}\right]
\eqno(4.9)
$$
for the two lowest lying levels, with
$$
S_0={1\over 6}\;\tilde \eps_d\omega b^2\quad.
\eqno(4.10)
$$
The WKB exponential factor $S_0$ is just the ``kink energy'' of a
defect with stiffness $\tilde\epsilon_d$
which moves between the two sites as a function of $z$ in the
path integral (1.2). Upon identifying the splitting in (4.9) with $t$ in
Eq.~(4.3), we find
$$
t=2\sqrt{{3\over 2\pi}}\;
\omega(TS_0)^{1/2}
e^{-S_0/T}
\quad.
\eqno(4.11{\rm a})
$$

To a zeroeth approximation, we have $u_0\equiv\eps_d$, the energy of the
centered interstitial computed in Sec.~3. In principle, $u_0$ should be
corrected by the zero point energy of a two-dimensional quantum oscillator,
i.e., by twice the first term of Eq.~(4.9). This represents the
entropy of harmonic  fluctuations of the defect.
There is also, however, a
{\it negative} contribution of this form from the entropy of the flux line
lattice
itself [25] which should have approximately the same magnitude. We assume
for simplicity here that
these two contributions simply cancel.

It is convenient now to set
$$
u_0=\alpha\Delta\ep\quad,
\eqno(4.11{\rm b})
$$
where according to Table 3, $\alpha\approx 70$. Upon using Eqs.~(4.11) and
(4.10), we find that the condition $E_0(T_d)=0$ takes the form
$$
\alpha\left({S_0\over T_d}\right)^{1/2}=
32\sqrt{{3\over 2\pi}}\;e^{-S_0/T_d}\quad,
\eqno(4.12)
$$
which  is solved by $S_0/T_d=0.085$. The assumption
$\tilde \eps_d\approx\tel$, then leads to
$$
T_d=0.30\sqrt{\eps_0\tel}
\;a_0\quad,
\eqno(4.13)
$$
i.e., Eq. (2.24) with $c_3=0.30$. The substitution $c_L=0.15-0.30$ in
Eq.~(2.11), however, shows that the melting temperature $T_m$ is
significantly (about an order of magnitude) {\it smaller} than our
estimate of $T_d$ in the regime $B_{{\rm c1}}<B<B_\times$. Evaluation
of the {\it high} field formula Eq.~(2.26) at $B=B_\times$ is only slightly
more encouraging:
Using Table~3 for the centered interstitial energy leads to
$T_d=0.079\;\eps_0d_0$, while the Lindemann criteria at $B=B_\times$ gives
$T_m=0.02-0.09\;\eps_0d_0\sim T_d$ [44]. We conclude that the supersolid
probably only
exists for $B>B_\times$, as indicated in Fig.~5.

Of course, our estimate for $S_0/T_d$ is so small that it casts doubt on the
validity of the tight-binding model and the WKB approximation, which are
only strictly correct at low
temperatures. More accurate band structure estimates of $T_d$ for $B<B_\times$
(including $B<B_{{\rm c1}}$) would certainly be of interest.

\noindent
{\caps 4.2 Nature of the Transition at $T_d$}

As discussed above, a transition to a supersolid below the equilibrium
melting temperature of the flux crystal becomes possible when $B>B_\times$.
Once
the defects proliferate, interactions between them become important, as in the
closely related problem of vortex penetration just above H$_{{\rm c1}}$. To
model this process, we use a continuum coherent state path integral
representation of the partition function, similar
to one used for flux lines near H$_{{\rm c1}}$ [7]. The
details of the lattice of preferred sites are neglected,
although these could easily
be taken into account if necessary. The grand canonical partition function
describing the interstitial degrees of freedom then reads
$$
\CZ_{gr}=
\int\CD\psi_i\CD\psi^*e^{-S[\psi,\psi^*]}
\eqno(4.14)
$$
with
$$
S[\psi,\psi^*]=
\int d^2r\;dz
\left[\psi_i^*\left(\partial_z-
{T\over 2\te}\nabla_\perp^2\right)
\psi_i+r|\psi_i|^2+
u|\psi_i|^4+
\cdots\right]\quad.
\eqno(4.15)
$$
Here $\psi_i(\vr,z)$ and $\psi_i^*(\vr,z)$ represent interstitial line
creation and annihilation operators, and the $r\propto (T_d-T)$ is the defect
chemical potential. The areal density of interstitials $n_i$ is given by
$$
n_i=|\psi_i(r,z)|^2\quad,
\eqno(4.16)
$$
and the couplings $u$ and $v$ represent the effect of interactions.

The nature of the transition to the supersolid which occurs with decreasing
$r$ depends crucially on the sign of the quartic coupling in Eq.~(4.15).
If $u>0$, then a continuous transition results, and $n_d\propto(T-T_d)$
up to logarithmic corrections. Both the coefficient and the logarithmic
correction can
be calculated as in [7]. If $u<0$, the transition to the supersolid becomes
first order. A first-order transition is possible because the
microscopic two-body interaction  between
centered interstitials (represented by $u|\psi(\vr,z)|^4$ in Eq.~(4.15)) is
{\it attractive}, at least  at large distances; see Section~3.
The unbinding temperature $T_d<T_m$ is probably too low to allow for a
significant thermal renormalization of $u$, which could in principle be
driven positive by entropic effects. The first-order transition described
by (4.15) with $u$ negative is discussed in Ref.~45.

A finite density of proliferating defects is detectable, at least in
principle, via a neutron diffraction experiment which precisely determines the
temperature dependent magnitude  of the six smallest reciprocal lattice vector
$\{\vG(T)\}$ of the
Abrikosov flux array for fixed magnetic field $\vB$. Note that $\vB$ will
also be, in general,
temperature-dependent for fixed external field $\vH$.
The number of unit cells per unit area $n_c$ is related to $G$ according to
$$
n_c(T)=\sqrt3\; G^2(T)/8\pi^2\quad.
\eqno(4.17)
$$
If vacancies or interstitials only exist in small closed loop pairs, as in
Fig.~1b, $n_c$ must be {\it exactly} equal to the areal density of
vortices, $n_c=n_0=B/\phi_0$. Above the proliferation temperature $T_d$,
interstitials dominate over vacancies, and we have
$$
\eqalignno{
n_i(T)&=\la|\psi_i(\vr,z)|^2\ra\cr
&=n_0-n_c(T)>0\quad.&(4.18)
\cr}
$$
A plot of $n_c(T)/n_0$ is thus a direct measure of the density of defects, with
$n_c(T)/n_0\equiv 1$ for $T<T_d$, and $n_c(T)/n_0<1$ when $T>T_d$.

A nonzero defect order parameter
$\la\psi_i(r,z)\ra$ necessarily implies that the boson order parameter (1.6)
is nonvanishing, because wandering vacancies or interstitial defects catalyze
enhanced entanglement of the underlying vortex crystal. The vortices are thus
simultaneously crystalline and entangled when $T_d<T<T_m$, as discussed in the
Introduction. Once the equilibrium concentration of one type of defect is
nonzero, {\it all} defects will proliferate in at least small
concentrations. Consider, in particular, the process shown in Fig.~1b,
the creation of vacancy-interstitial pairs. This can be modeled by adding
terms to the action (4.15) as follows:
$$
S\rightarrow S+\int d^2rdz\left[
\psi_v^*\left(\partial_z-
{T\over 2\te}\nabla_\perp^2\right)
\psi_v+r_v|\psi_v|^2+g(\psi_i
\psi_v+\psi_i^*\psi_v^*)\right]
\quad.
\eqno(4.19)
$$
The first two terms are the vacancy propagator and chemical potential, while
the last allows pair creation and destruction with probability
proportional to $g$ [46]. Because the vacancies are
unfavorable relative to interstitials, $r_v$ will remain positive just
above $T_d$. We see, however, that a nonzero $\la\psi_i(\vr,z)\ra$ acts like an
ordering field on $\psi_v(r,z)$, i.e., $\la\psi_v(\vr,z)\ra\not= 0$
for $T>T_d$. In more physical terms, vacancies can appear
because their unfavorable energy in isolation is compensated by nearby
thermally created interstitials.

We also mention an exotic type of supersolid which is possible at least in
principle: Suppose that the split interstitial, or some similar defect had
the lowest energy. The three distinct orientations of this defect within
its hexagonal cell represent an internal degree of freedom for the
corresponding ``boson'' world lines. Should such ``ribbon-like''
defects proliferate in the solid, the resulting fluid of lines  would be a
``quantum
rotator'' liquid,  with the same potential broken symmetries as a quantum
three-state Potts model when interline interactions are taken into account.

\noindent
{\caps 4.3 Nature of the Transition at $T_m$}

It is important to distinguish between Type~I and Type~II melting
into a liquid phase, according to whether vacancy and interstitial defects
have already proliferated---see Fig.~4. Consider the standard Landau
argument for a first-order transition starting from the flux liquid
state: Provided fluctuations suppress crystallization well below the
mean field H$_{{\rm c2}}$, one can expand the local BCS condensate
density in Fourier components of the incipient density wave,
$$
\la|\psi_{BCS}(\vr,z)|^2\ra=\rho_0
+\sum_{\vG}\rho_{\vG}e^{i\vG\cdot\vr}\quad,
\eqno(4.20)
$$
where the $\{\vG\}$ are reciprocal lattice vectors in a plane perpendicular
to $\hat z$. The free-energy difference $\delta{\cal F}$ between the liquid and
crystalline phases can then be expressed as a Taylor series in the order
parameters $\{\rho_{\vG}\}$,
$$
\delta{\cal F}={1\over 2}\;r_G\sum_{j=1}^6
|\rho_{\vG_j}|^2+w
\sum_{\vG_i+\vG_j+\vG_k=0}
\rho_{\vG_i}\rho_{\vG_j}
\rho_{\vG_k}+
\cdots,
\eqno(4.21)
$$
where we have included only the first ring of six smallest $\vG$'s. The
crucial element is the third-order term allowed by the symmetry of a
triangular lattice, which leads to a first-order transition within this mean
field theory [2]. The magnitude of the smallest $G$'s in Eq.~(4.20) would
be completely determined by the magnetic field for Type~I melting, but would
depend on the incipient density of vacancies or interstitials in
the Type~II case, as discussed in Sec.~4.2.

Tesanovic [42] has suggested that when the freezing transition approaches
$H_{c2}$, as it will at low temperatures or if the fluctuations are weak,
the transition will be to a charge density wave state with a wavevector
not simply related to the magnetic field. Such a state is similar to the
``supersolid'' phase discussed here, and presumably evolves continuously into
it as the melting field falls well below $H_{c2}$. The line of supersolid
phase transitions may also be related to the solid-phase transition
suggested by Glazman and Koshelev [24], above which phase coherence is lost
in the $z$ direction. See Sec.~4.4.

Note that generically, if a melting transition is weakly first order,
it is likely to be to an incommensurate solid phase as the
wave-vector-dependent $r_G$ in Eq.~(4.21) will in general  not
have its minimum at any particularly simple wavevector.

The distinction between Type I and Type II behavior also affects dislocation
mediated melting theories, which start from the ordered phase just below
$T_m$. Reference 8, for example, studies effects of dislocation loops
which are confined to the plane spanned by their Burger's vector and the
$z$-axis. This restriction implicitly assumes Type~I melting, i.e., that
no vacancies or interstitials are present at long wavelengths. The absence of
vacancy or interstitial lines means that only ``glide'' motions of
dislocation lines are allowed along the time-like $z$ axis, which is
equivalent to a planarity restriction for the three-dimensional vortex
loops. A small concentration of proliferating vacancy or interstitial lines
would allow ``climb-like'' distortions of a vortex loop, as the loop
absorbs or emits these defects. Type~II dislocation mediated melting
would thus involve arbitrary nonplanar dislocation loop configurations, in
contrast to the planar loops associated with Type~I melting.

\noindent
{\caps 4.4 Transport Properties}

Finally, we compare briefly the resistive properties of the supersolid,
crystal and vortex fluid phases.

In the absence of pinning by random inpurities, all  the above phases
dissipate currents perpendicular to the macroscopic magnetic field
since the vortex lines can move freely provided---in the solid--the
motion is uniform. In contrast, the vortex crystal is a linear
superconductor for currents parallel to the $z$ direction. Concommitantly,
it can screen additional magnetic fields normal to the $z$ direction.
This will not occur in a semi-infinite
system with a planar surface since the magnetic fields can just
rotate. But in a cylinder with the vortex parallel to the axis, a small
additional azimuthal magnetic field will be expelled, decaying
exponentially in a layer with  thickness given by an  effective penetration
length~[48].

Even in the crystal phase, {\it finite} currents parallel to $\hat z$ will
be dissipated by nonlinear nucleation of vorticity normal to $\hat z$;
the details of this process in bulk samples have not, to our knowledge, been
analyzed.

It is interesting to note that while the vortex crystal is very anisotropic
in response to uniform currents, it appears much more isotropic in its
linear confinement of magnetic monopoles as discussed in the Introduction.

Both the vortex supersolid and vortex fluid phases will respond like
normal metals to currents parallel to $\hat z$, although there will
be additional nonlinear dispersion and nonlocal effects [49].
The linear resistivity may also be extremely anisotropic if the vortex
lines are rather straight. In the supersolid phase, the dissipation
will be dominated by the fluid of interstitial (or vacancy) lines
while the underlying lattice will not move in response to a small
current in the $z$ direction.

Pinning by
random impurities,
oxygen vacancies or other defects strongly affects the resistive
properties of the mixed state of superconductors. If the pinning
is very weak, the thermodynamic properties will be little affected
but the long-range translational order of both the crystal and supersolid
phases
will be destroyed on long distances, resulting in a large but finite
positional correlation length. The crystal phase may  be replaced
in its entirety by a truly superconducting vortex glass
phase [12] with vanishing  linear resistivity. In contrast, the vortex
fluid phase is not much affected  by weak pinning. Because  the
defects are  fluid-like
in the supersolid phase, they also will not be strongly affected by
weak pinning, provided they are sufficiently dense [50]. The
resulting phase will then still be metallic with
nonzero resistivity. Thus, it appears likely that the vortex-glass
phase transition for weak pinning should  occur at what was the defect
proliferation transition, $T_d$, in the pure system. If the vortex lattice
melting is Type $I$, then the putative vortex-glass transition should
probably occur at the $T_m$ of the pure system.

The nonlinear response of the weakly pinned supersolid will involve
motion of  both the underlying lattice, including plastic flow
involving dislocations, as well as vacancies and interstitials. Thus, even if
the
defect lines themselves are pinned, and the supersolid  is in a vortex glass
phase, the nonlinear response might distinguish it from a weakly
pinned crystal.

In the presence of sufficiently strong pinning, both equilibrium phase
transitions---defect
proliferation and melting---will  be destroyed.
The vortex glass transition, if it exists at finite
temperatures, will then become second order with the
resistance vanishing continuously as the temperature is decreased.
In this case, the distinctions between the low temperature solid
phases will disappear as the extent of any crystaline order will be
very short range. There are, however, several caveats.
Firstly, it is likely that hexatic bond-orientational order can persist
out to much longer distances than positional order [51], and perhaps even to
infinite distances so that distinct hexatic and vortex glass transitions
could occur. Another possibility is that more than one type of vortex
glass phase could exist, one with residual shear elasticity, as hypothesized
by Feigel'man \etal~[53] and one without shear elasticity as discussed
by Fisher, Fisher and Huse~[12].
Which of these phases actually can exist even in principle is still
very unclear. This complex set of possibilities illustrates the subtleties
once random point pinning is taken into account.

As mentioned earlier, the basic phase boundaries may well, except
for strong pinning, be determined by the properties of the pure
vortex systems analyzed in this paper. Thus, a more quantitative
analysis of the phase diagram in the pure system is certainly merited.

\vfill
\noindent
{\bf Acknowledgements}
We have benefited from conversations with D.J. Bishop, V. Elser, D.A. Huse, and
C.M. Murray. This work was supported by National Science Foundation,
through Grants No. DMR91--15491 and DMR91--06237
and through the Harvard Materials
Research Laboratory.
The work of Erwin Frey has been supported by the Deutsche
Forschungsgemeinschaft  (DFG) under Contracts No. Fr. 850/2-1,2.
\eject

\noindent
{\caps Appendix A: Calculation of the Dynamic Matrix in the Limit
$\lambda_\perp\to\infty$.}

In this Appendix we describe the details of the calculation of the
dynamic matrix. Starting from Eq.~(3.16) the central quantity to
calculate is
$$
I({\vec x}/\lambda)=\sum_l
e^{-i{\vec q}\cdot{\vec x}(l)}K_0(|{\vec x}(l)-{\vec x}|/
\lambda)-K_0(|{\vec x}|/\lambda)\;,
\eqno({\rm A1})
$$
where we set $\lambda_\perp\equiv\lambda$ in what follows.
Upon using the integral representation of the Bessel function $K_0$
$$
K_0(x)={1\over2}\int_0^\infty d\tau\,\tau^{-1}e^{-\tau}\exp
\left[-{x^2\over4\tau\lambda^2}\right]\;,
\eqno({\rm A2})
$$
and Ewald's generalized theta-function transform [39]
$$
\sum_l e^{-i{\vec q}\cdot{\vec x}(l)}\exp[-|x(l)-x|^2t]=
{\pi\over a_ct}\sum_{\vec G} e^{-i({\vec q}+{\vec G})
\cdot{\vec x}}\exp[-|{\vec q}+{\vec G}|^2/4t]\;,
\eqno({\rm A3})
$$
where $a_c$ is the volume of the unit cell, one can rewrite the first
term in Eq.~(A1) as
$$
\eqalignno{
I_1({\vec x}/\lambda)\,&=
\sum_l e^{-i{\vec q}\cdot{\vec x}(l)}K_0(|{\vec x}(l)-{\vec x}|/\lambda^2)\cr
&={1\over2}\int_0^\infty d\tau\,\tau^{-1}e^{-\tau}
\sum_l e^{-i{\vec q}\cdot{\vec x}(l)}\exp\left[-{|{\vec x}(l)-{\vec x}|^2
\over 4\tau\lambda^2}\right]\cr
&={2\pi\over a_c}\,\lambda^2\int_0^\infty d\tau\,e^{-\tau}\sum_{\vec G}
e^{-i({\vec q}+{\vec G})\cdot{\vec x}}
\exp[-\tau\lambda^2|{\vec q}+{\vec G}|^2]\;.&({\rm A4})\cr}
$$
Now we split the $\tau$-integration by introducing an arbitrary
Ewald split-parameter $\epsilon$
$$
\eqalignno{
I_1({\vec x}/\lambda)=\,&
{1\over2}\int_0^\epsilon d\tau\,\tau^{-1}e^{-\tau}
\sum_l e^{-i{\vec q}\cdot{\vec x}(l)}\exp\left[-{|x(l)-x|^2
\over4\tau\lambda^2}\right]\cr
&+{2\pi\over a_c}\,\lambda^2\sum_{\vec G}
e^{-i({\vec q}+{\vec G})\cdot{\vec x}}
{e^{-\epsilon(\lambda^2|{\vec q}+{\vec G}|^2+1)}
\over1+\lambda^2|{\vec q}_{\vec G}|^2}\;,
&({\rm A5})\cr}
$$
where we have already performed the $\tau$-integral in the second term.

We are interested in the limit of large London penetration depth $\lambda$.
Upon choosing $\epsilon\sim\lambda^{-2}$ we take the limit $\lambda\to\infty$
with $\epsilon\lambda^2$ staying finite. Then one gets for the second term
in Eq.~(A.5)
$$
{2\pi\over a_c}\,\lambda^2\sum_{\vec G} e^{-i({\vec q}+{\vec G})\cdot{\vec x}}
{e^{-\epsilon\lambda^2|{\vec q}+{\vec G}|^2}\over1+\lambda^2|{\vec
q}_{\vec G}|^2}\;.
\eqno({\rm A6})
$$
For the first term we make the substitution
$ y = |{\vec x}(l)-{\vec x}|^2 / (4\tau\lambda^2)$ and get
$$
{1\over2}\int_a^\infty {dy\over y}\,\sum_l\exp
\left[-{|{\vec x}(l)-{\vec x}|^2\over4y\lambda^2}\right]
e^{-y}e^{-i{\vec q}\cdot{\vec x}(l)}\;.
\eqno({\rm A7})
$$
In the limit $\lambda\to\infty$ we get
$\exp\left[-{|{\vec x}(l)-{\vec x}|^2\over4y\lambda^2}\right]\to1$
since $y$ is bounded from below by $a={1\over4\epsilon\lambda^2}\,
|{\vec x}(l)-{\vec x}|^2$. Hence the first term can be written in terms of the
exponential integral function which is defined by
$$
E_1(x)=-Ei(-x)=\int_x^\infty {dy\over y}\,e^{-y}\;.
\eqno({\rm A8})
$$
In summary one gets
$$
\eqalignno{
\lim_{\lambda\to\infty} I_1({\vec x}/\lambda)=\,&
{1\over2}\sum_l e^{-i{\vec q}\cdot{\vec x}(l)}E_1
\left(|{\vec x}(l)-{\vec x}|^2\over4\epsilon\lambda^2\right)\cr
&+
{2\pi\over a_c}\,\lambda^2\sum_{\vec G} e^{-i({\vec q}+{\vec G})\cdot{\vec x}}
{e^{-\epsilon\lambda^2|{\vec q}+{\vec G}|^2}\over|{\vec q}+{\vec G}|^2}\;.
&({\rm A9})\cr}
$$
In order to get $I(|{\vec x}|/\lambda)$ one has to subtract
$$
K_0(x/\lambda)={1\over2}\int_0^\infty
d\tau\,\tau^{-1}e^{-\tau}e^{-x^2/4\lambda\tau}\approx
- -\ln(x/2\lambda)\;
\eqno({\rm A10})
$$
from $I_1(|{\vec x}|/\lambda)$ in the limit $\lambda \to \infty$.
Therefrom one can now calculate $C_{\alpha\beta}$ and expand in
powers of the wave vector ${\vec q}$. The result is
$$
C_{\alpha\beta}({\vec q})=
{\phi_0^2\over8\pi^2\lambda^2}\,
\left\{2\pi{q^\alpha q^\beta\over q^2}+{a_c\over8}\,
[\delta^{\alpha\beta}q^2-2q^\alpha q^\beta]\right\}\;.
\eqno({\rm A11})
$$
where we have taken $1/(4\epsilon\lambda^2)=\pi/a_c$ for
the Ewald separation parameter.
\bigskip

\noindent
{\caps Appendix B: Variational Calculation of the Vacancy Formation Energy}

In this appendix we give the details of the variational calculation
for the vacancy formation energy.
Upon using the long wavelength approximation (3.18) for
$C^{\alpha\beta}(\vq)$, one finds three contributions to $E_{\rm VR}^{(2)}$,
the vacancy relaxation energy at constant lattice spacing,
$$
E_{\rm VR}^{(2)} ( \{ \vu (\ell) \} )  = 2\epsilon_0 (I_1 + I_2 + I_3)
\eqno({\rm B1})
$$
with
$$
\eqalignno{
I_1\,&=
{1\over2}\int{d^2q\over(2\pi)^2}\,(i)^2\,
{q^\alpha\over q^2}\,{-q^\beta\over q^2}\,f^2(q)
\left\{2\pi\,{q^\alpha q^\beta\over q^2}+{a_c\over8}\,
(\delta^{\alpha\beta}q^2-2q^\alpha q^\beta)\right\}\cr
&=
{1\over2}\int_0^{k_d} dq\,q^{-1}f^2(q)-{a_c\over32\pi}\,
\int_0^{k_d} dq\,qf^2(q)\;,
&({\rm B2})\cr}
$$
$$
\eqalignno{
I_2\,
&=-\int_q i\left({2\pi\over q^2}-{a_c\over2}\right)
q^\alpha i\,{-q^\alpha\over q^2}\,f(q)\cr
&=-\int_0^{k_d} dq\,q^{-1}f(q)+{a_c\over4\pi}
\int_0^{k_d} dq\,qf(q)\;,
&({\rm B3})\cr}
$$
and
$$
\eqalignno{
I_3\,
&=-{1\over2a_c}\int_q\int_k(-1)
\left\{{2\pi(q-k)^\alpha(q-k)^\beta\over({\vec q}-{\vec k})^2}+{a_c\over8}\,
\left(\delta^{\alpha\beta}({\vec q}-{\vec
k})^2-2(q-k)^\alpha(q-k)^\beta\right)\right\}\cr
&\qquad\times(i)^2a^2_c\,{k^\alpha\over k^2}\,f(k)\,{-q^\beta\over
q^2}\,f(q)\cr
&={a_c\over2}\int_q\int_k f(q)f(k)
\left[2\pi{({\vec q}\cdot{\vec k}-k^2)(q^2-{\vec q}\cdot{\vec k})
\over(q^2-2{\vec q}\cdot{\vec k}+k^2)q^2k^2}+
{a_c\over8}\,({\vec q}\cdot{\vec k})(q^2-2{\vec q}\cdot{\vec
k}+k^2)\,{1\over q^2k^2}\right.\cr
&\qquad-\left.
{a_c\over4}\,({\vec q}\cdot{\vec k}-k^2)(q^2-{\vec q}\cdot{\vec
k})\,{1\over q^2k^2}\right]\;,
&({\rm B4})\cr}
$$
where $f(q)=f(1+cq+dq^2)$ and we have approximated the hexagonal Brillouin
zone by a circle of equal area with radius $k_d=\sqrt{4\pi\over a_c}$.

The third contribution can be split into $I_3=I_{3a}+I_{3b}+I_{3c}$
where
$$
I_{3a}=
a_c\pi\int_0^{k_d} q\,dq\int_0^{k_d} k\,dk\,{2\pi\over(2\pi)^4}
\int_{-\pi}^\pi d\varphi\,f(p)f(k)\,
{(qk\cos\varphi-k^2)(q^2-qk\cos\varphi)\over
(q^2-2qk\cos\varphi+k^2)q^2k^2}\;.
\eqno({\rm B5})
$$
Upon using
$$
F(q_1,q_2)=
\int_{-\pi}^\pi d\varphi\,
{(q_2^2-q_1q_2\cos\varphi)(-q_1^2-q_1q_2\cos\varphi)\over
q_1^2+q_2^2-2q_1q_2\cos\varphi}=\cases{
-\pi q_2^2& for $q_1>q_2$\cr
-\pi q_1^2& for $q_1<q_2$\cr}
\eqno({\rm B6})
$$
one arrives at
$$
\eqalignno{
I_{3a}\,
&=-{a_c\over8\pi}\int_0^{k_d}dq\int_0^{k_d}\,{f(q)f(k)\over qk}\,
\cases{k^2&for $k<q$\cr q^2&for $k>q$\cr}\cr
&=-{a_c\over8\pi}\int_0^{k_d}dq
\left(\int_0^q{k\over q}\,f(q)f(k)\,dk+
\int_q^{k_d}{q\over k}\,f(q)f(k)\,dk\right)\;,
&({\rm B7})\cr
I_{3b}\,
&={a_c^2\over16}\int_0^{k_d}q\,dq\int_0^{k_d}k\,dk\,{1\over(2\pi)^3}
\int_{-\pi}^\pi d\varphi\,{\cos\varphi\over qk}\,f(p)f(k)
(q^2-2qk\cos\varphi+k^2)\cr
&=-{a_c^2\over16(2\pi)^2}\int_0^{k_d}dq\int_0^{k_d}dk\,f(q)f(k)qk\;,
&({\rm B8})\cr
I_{3c}\,
&=-{a_c^2\over8}
\int_0^{k_d}q\,dq\int_0^{k_d}k\,dk\,
{1\over(2\pi)^3}\,{1\over q^2k^2}\,
\int_{-\pi}^\pi d\varphi\,f(p)f(k)
(qk\cos\varphi-k^2)(q^2-qk\cos\varphi)\cr
&={3a_c^2\over 16(2\pi)^2}
\int_0^{k_d}dq\int_0^{k_d}dk\,f(q)f(k)qk\;.
&({\rm B9})\cr}
$$
All these integrals diverge at small momenta. This is due to the fact that we
have taken the penetration depth $\lambda\equiv\lambda_\perp$
to be infinite. Upon reintroducing
the lower cutoff $\lambda^{-1}$ the leading contribution
(which diverges in the limit $\lambda\to\infty$) is found to be
$$
\eqalignno{
I_1\,&={f^2\over2}\,\ln(k_d\lambda)+\CO(1)\;,
&({\rm B10})\cr
I_2\,&=-f\,\ln(k_d\lambda)+\CO(1)\;,
&({\rm B11})\cr
I_3\,&=\CO(1,(\ln\lambda)/\lambda^2)\;.
&({\rm B12})\cr}
$$
Hence one finds
$$
E_{\rm VR}^{(2)}=
{\phi^2_0\over8\pi^2\lambda^2}
\left[{f^2\over2}\,\ln(k_d\lambda)-f\,\ln(k_d\lambda)
+ {1\over 2} f-{3f^2\over 16}\right]\;.
\eqno({\rm B13})
$$
Since we are interested in the limit $\lambda\rightarrow\infty$ we must choose
$f=1$ in
order to cancel the $\ln(k_d\lambda)$ divergences in the final expression for
the
vacancy-free energy.

%
Now we have to look at the subleading terms. We set $f=1$ and do variational
calculation for $c$ and $d$. One gets
$$
c={4\over7k_d}\approx 0.161\sqrt{a_c}\;,
\eqno({\rm B14})
$$
$$
d=-{10\over7k_d^2}\approx-0.114 a_c\;.
\eqno({\rm B15})
$$
The result for the vacancy relaxation energy at constant lattice
spacing becomes
$$
E_{\rm VR}^{(2)}=
{\phi^2_0\over 32 \pi^2\lambda^2}\,
\left[\ln\left({1 \over k_d^2\lambda^2} \right)+{265\over252}\right]\;,
\eqno({\rm B16})
$$
which corresponds to a free energy of
$$
G_D=\left(
{\phi_0\over 8\pi\lambda}\right)^2
0.789\;.
\eqno({\rm B17})
$$

\vfill\eject

\centerline{\caps REFERENCES}

\item{1.} See, e.g., the reviews in {\it Phenomenology and
Applications of High-Temperature Superconductors}, edited by K. Bedell
{\it et al.} (Addison-Wesley, New York, 1991).
\item{2.} E. Brezin, D.R. Nelson, and A. Thiaville, Phys. Rev. B{\it 31}, 7124
(1985).
\item{3.} M. Charalambous, J. Chaussy and P. Lejay,   Phys. Rev. B{\it 45},
509 (1992).
\item{4.} H. Safar, P.L. Gammel, D.A. Huse, D.J. Bishop, J.P. Rice, and D.M.
Ginsberg, Phys. Rev. Lett. {\it 69}, 824 (1992).
\item{5.} K.W. Kwok, S. Fleshler, U. Welp, V.M. Vinokur, J. Downey,
G.W. Crabtree and M.M. Miller, Phys. Rev. Lett. {\it 69}, 3370 (1992).
\item{6.} A.I. Larkin and Y.M. Ovchinnikov, J. Low Temp. Phys. {\it 34},
409 (1979).
\item{7.} D.R. Nelson and H.S. Seung, Phys. Rev. B{\it 39}, 9153 (1989).
\item{8.} M.C. Marchetti and D.R. Nelson, Phys. Rev. B{\it 42}, 9938 (1990).
\item{9.} F.R.N. Nabarro and A.T. Quintanilha, in {\it Dislocations in
Solids}, edited by F.R.N. Nabarro (North-Holland, Amsterdam, 1980), Vol. 5.
\item{10.} N.W. Ashcroft and N.D. Mermin, {\it Solid State Physics}
(Saunders College, Philadelphia 1976), Chapter 30.
\item{11.} The dislocation loop in Fig.~1a differs from the
vacancy-interstitial pair in Fig.~1b in that it relaxes {\it shear
stresses}
rather than relaxing a perpendicular current directly.
\item{12.} D.S. Fisher, M.P.A. Fisher, and D.A. Huse, Phys. Rev. B{\it
43}, 130 (1991).
\item{13.} M. Feigel'man, V.B. Geshkenbein, and A.I. Larkin, Physica
C{\it 167}, 177 (1990).
\item{14.} V.M. Vinokur, P.H. Kes, and A.E. Koshelev, Physica C{\it 168},
29 (1990).
\item{15.} J.H. Hetherton, Phys. Rev. {\it 176}, 231 (1968).
\item{16.} A.F. Andreev and I.M. Lifshitz, Sov. Phys.  JETP {\it 29},
1107 (1969).
\item{17.} G. Chester, Phys. Rev. A{\it 2}, 256 (1970).
\item{18.} A.J. Leggett, Phys. Rev. Lett. {\it 25}, 1543 (1970).
\item{19.} I.E. Dzyaloshinskii, P.S. Kondratenko, and V.S. Levchenkov,
Sov. Phys. JETP {\it 35}, 823, 1213 (1972).
\item{20.} K. Liu and M.E. Fisher, J. Low Temp. Phys. {\it 10}, 655 (1973).
\item{21.} M.P.A. Fisher and D.H. Lee, Phys. Rev. B{\it 39}, 2756 (1989).
\item{22.} D.R. Nelson, in Ref.~1.
\item{23.} A.F. Andreev, ``Quantum Crystals,'' in {\it Progress in
Low-Temperature
Physics}, Vol. VIII, edited by D.G. Brewer (North Holland, Amsterdam, 1982).
\item{24.} L.I. Glazman and A.E. Koshelev, Phys. Rev. B{\it 43}, 2835 (1991).
\item{25.} (a) D.S. Fisher, B.I. Halperin and R. Morf, Phys. Rev. B{\it 20},
4692 (1979); This work is revised and extended in (b)
E. Cockayne and V. Elser, Phys. Rev. B{\it 43}, 623 (1991).
\item{26.} Several interstitials are in fact evident in the triangulation of
decorated vortices in BSCCO of Fig. 2 in P. Gammel, D.J. Bishop and
C.M. Murray, Phys. Rev. Lett. {\it 64}, 2312 (1990).
These interstitials, however, may be an
artifact of overdecoration (C.M. Murray, private communication). See also C.A.
Bolle {\it et al.}, Phys. Rev. Lett. {\it 66}, 112 (1991) for chains of
interstitials which appear in magnetic fields tilted away from the $c$ axis.
One
possible explanation for the chains is a highly anisotropic interaction
potential
between interstitials.
\item{27.} A. Houghton, R.A. Pelcovits, and A. Sudbo, Phys. Rev.
B{\it 40}, 6763 (1989), and Refs. 13, 14, 25, and 24.
\item{28.} D.S. Fisher in Ref. 1 and Ref. 25.
\item{29.} E.H. Brandt, Phys. Rev. B{\it 34}, 6514 (1986); Phys. Rev.
Lett. {\it 63},
1106 (1989).
\item{30.} See Appendix B of D.R. Nelson and V. Vinokur, Phys. REv. B (in
press).
\item{31.} E.H. Brandt, Phys. Rev. Lett. {\it 69}, 1105 (1992).
\item{32.} R.P. Feynman and A.R. Hibbs, {\it Path Integrals and Quantum
Mechanics}, (McGraw-Hill, New York, 1965); R.P. Feynman, {\it Statistical
Mechanics} (W.A. Benjamin, Reading, MA, 1972).
\item{33.} See Ref. 32 and Appendix C of Ref. 30.
\item{34.} D.S. Fisher, Phys. Rev. B{\it 22}, 1190 (1980).
\item{35.} D.R. Nelson, J. Stat. Phys. {\it 57}, 511 (1989).
\item{36.} A.L. Fetter, Phys. Rev.\ {\it 147}, 153 (1966); The values
given for $A_4$
and $A_6$ by Fetter are corrected by the values given in the main text.
We attribute the
small difference to inaccuracies in the evaluation of the exponential integral
function.
\item{37.} The actual stable vacancy configuration does not have this six-fold
symmetry. The vacancy configuration with the six-fold symmetry is a
saddle point. See
also section 3.5.
\item{38.} A.L. Fetter, P.C. Hohenberg and P. Pincus,
Phys. Rev.\ {\it 147}, 140 (1966).
\item{39.} P.P. Ewald, Ann. Phys.\ (Leipzig)
{\it 54} 519 (1917); {\it 54}, 57 (1917); {\it 64} 253 (1921).
\item{40.} M.P. Allen and D.J. Tildesley, {\it Computer Simulations of
Liquids}, (Clarendon, Oxford, 1990).
\item{41.} We have used the polynomial and rational approximations 5.1.53 and
5.1.56
in M. Abramowitz, and I.A. Stegun, {\it Handbook of Mathematical Functions},
Dover
Publ., Inc., New York (1965), page 231.
\item{42.} The error for the total energy of the bigger system is
larger because we have
used a square cell with a length to height ratio of ${13\over 15}$ instead of a
hexagonal cell (see above). The error implied by this becomes larger for larger
systems.
\item{43.} See, e.g., S. Coleman, {\it Aspects of Symmetry} (Cambridge
University
Press, 1987), Chapter 7; see also J.F. Currie {\it et al.}, Phys. Rev. B{\it
22},
477 (1980).
\item{44.} The two-dimensional formula Eq. (2.14) gives $T_m=0.01\;c_0d_0$,
which
is also less than $T_d$.
\item{45.} R.D. Kamien and D.R. Nelson, J. Stat. Phys. {\it 71}, 23 (1993); see
Appendix B.
\item{46.} For a related treatment of hairpins in polymer nematics, see
R.D. Kamien,
P. Le Doussal and D.R. Nelson, Phys. Rev. A{\it 45}, 8727 (1992).
\item{47.} Z. Tesanovic, Johns Hopkins University preprint.
\item{48.} Similar issues are discussed in E.H. Brandt, J. Low. Temp. Phys.
{\it 42}, 557 (1981); ibid {\it 44}, 33 (1981); ibid {\it 44}, 59 (1981).
\item{49.} D. Huse and Majumdar, AT\&T Bell Laboratories preprint; see also
C.M. Marchetti and D.R. Nelson, Physica C{\it 174}, 40 (1991).
\item{50.} D.R. Nelson and P. Le Doussal, Phys. Rev. B{\it 42}, 10112 (1990).
\item{51.} E.M. Chudnovsky, Phys. Rev. B{\it 40}, 11355 (1989).
\item{52.} M.V. Feigel'man, V.B. Geshkenbein, A.I. Larkin and V.M. Vinokur,
Phys. Rev. Lett. {\it 63}, 2302 (1989).
\vfill\eject

\parindent=0pt
\centerline{\caps TABLE 1}
\medskip
\centerline{Estimates for Melting and Defect Unbinding Transitions}
\bigskip
\bigskip

\tabskip=1em plus 2em minus .5em
\halign to \hsize{\hfil#\hfil&\quad\hfil#\hfil&\quad\hfil#\hfil\cr
\noalign{\hrule}
\noalign{\vskip1pt}
\noalign{\hrule}
\noalign{\vskip6pt}
Regime&$T_m(B)$&$T_d(B)$\cr
\noalign{\vskip4pt}
\noalign{\hrule}
\noalign{\vskip6pt}
$B_\times\;\lot\;B$&$0.5 \eps_0d_0/8\pi\sqrt3$&${\rm
const.}\times\eps_0d_0/\ln(2\pi B/B_\times)$\cr\cr
$B_{c1} \ll B\;\lot\;B_\times $&$c_L^2\sqrt{\eps_0\tel}(\phi_0/B)^{1/2}$&$c_3
\sqrt{\eps_0\tel}(\phi_0/B)^{1/2}$\cr\cr
$B\;\lot\;B_{c1}$&$c_L^2\sqrt{\eps_0\tel}\lambda_\perp\left({B_{c1}\over
 B}\right)e^{-{1\over 2}(B_{c1}/B)^{1/2}}$&
${\rm const.}\times\sqrt{\eps_0\tel}\lambda_\perp\left({B_{\rm c1}\over
B}\right)
e^{-{1\over 2}(B_{c1}/B)^{1/2}}$\cr
\noalign{\vskip6pt}
\noalign{\hrule}
\noalign{\vskip1pt}
\noalign{\hrule}
\cr}
\vfill\eject
\centerline{\caps TABLE 2}
\medskip
\noindent Exact and numerically calculated energy per particle of the
perfect flux line lattice for different system sizes.
$N$ is the number of flux lines.
The energies are measured in units of $E_0 = {2 \eps_0}$.
\bigskip
\bigskip
\tabskip=1em plus 2em minus .5em
\halign to \hsize{\hfil#\hfil&\quad\hfil#\hfil&\quad\hfil#\hfil\cr
\noalign{\hrule}
\noalign{\vskip1pt}
\noalign{\hrule}
\noalign{\vskip6pt}
$N$&$E_{\rm exact}$&$E_{\rm num}$\cr
\noalign{\vskip4pt}
\noalign{\hrule}
\noalign{\vskip6pt}
$30$   &$0.85583194$  &$0.85583216$ \cr
$120$  &$1.20240554$  &$1.20240652$ \cr
$270$  &$1.40513809$  &$1.40514051$ \cr
$480$  &$1.54897913$  &$1.54898290$ \cr
$750$  &$1.66055090$  &$1.66055687$ \cr
$1080$ &$1.75171168$  &$1.75172015$ \cr
$1470$ &$1.82878705$  &$1.82879859$ \cr
\noalign{\vskip6pt}
\noalign{\hrule}
\noalign{\vskip1pt}
\noalign{\hrule}
\cr}
\vfill\eject
\centerline{\caps TABLE 3}
\medskip
\noindent
Defect energies at constant line density for the symmetric vacancy
(V6), ``squeezed''
vacancy (V2), edge interstitial (EI), centered interstitial (CI), split
edge interstitial (SEI), and split centered interstitial (SCI) configurations.
The results are obtained from a molecular dynamics type of calculation for
different system sizes with N flux lines.
The energies are measured in units of $E_0 = {2 \eps_0}$.
\bigskip
\bigskip
\tabskip=1em plus 2em minus .5em
\tabskip=1em plus 2em minus .5em
\halign to \hsize{\hfil#\hfil & \quad\hfil#\hfil & \quad\hfil#\hfil
& \quad\hfil#\hfil & \quad\hfil#\hfil & \quad\hfil#\hfil & \quad\hfil#\hfil\cr
\noalign{\hrule}
\noalign{\vskip1pt}
\noalign{\hrule}
\noalign{\vskip6pt}
$N$&$E_{\rm V6}$&$E_{\rm V2}$&$E_{\rm EI}$&$E_{\rm CI}$&$E_{\rm
SEI}$&$E_{\rm SCI}$\cr
\noalign{\vskip4pt}
\noalign{\hrule}
\noalign{\vskip6pt}
$30$   &$0.11857$    &$0.11394$   &$0.07156$   &$0.07218$   &$0.07156$
 &$0.07166$\cr
$120$  &$0.12204$    &$0.10892$   &$0.07358$   &$0.07274$   &$0.07359$
 &$0.07295$\cr
$270$  &$0.12381$    &$0.10825$   &$0.07392$   &$0.07293$   &$0.07392$
 &$0.07310$\cr
$480$  &$0.12416$    &$0.10783$   &$0.07403$   &$0.07299$   &$0.07402$
 &$0.07315$\cr
$750$  &$0.12429$    &$0.10781$   &$0.07408$   &$0.07304$   &$0.07405$
 &$0.07319$\cr
$1080$ &$0.12434$    &$0.10772$   &$0.07410$   &$0.07309$   &$0.07411$
 &$0.07324$\cr
$1470$ &$0.12437$    &$0.10780$   &$0.07412$   &$0.07307$   &$0.07411$
 &$0.07325$\cr
\noalign{\vskip6pt}
\noalign{\hrule}
\noalign{\vskip1pt}
\noalign{\hrule}
\cr}
\bigskip
\vfill\eject

\centerline{\bf Figure Captions:}
\bigskip
\bigskip
\noindent {\bf Fig.1a:}
Dislocation loop in a flux line solid. Dashed lines represent vortices
just behind the plane of the figure. Such loops lie in the plane spanned
by their Burgers vector and the $z$-axis. The orientation of the three
triangles is the same, showing that the loop has only a small effect on the
orientational order.
\bigskip
\noindent {\bf Fig.1b:}
Vacancy-interstitial pair in a flux line solid. Dashed lines represent vortices
just behind the plane of the figure. Unlike the dislocation loop in Fig.1a,
this loop is not constrained to lie in a single plane.
\bigskip
\noindent {\bf Fig.2:}
Vacancy line (thick dashed curve) meandering through a vortex crystal.
The full lines show the flux lines which are in the same plane
as the meandering vacancy. The dashed lines represent the
flux lines in the neighboring plane.
\bigskip
\noindent {\bf Fig.3:}
Lowest energy  contribution to the order parameter
correlation function on the solid phase  (1.4), this  inserts a flux head and
tail into
the crystalline vortex array. Dashed lines represent a row of vortices
slightly behind the plane of the page. A vacancy is created at ``time''
$z$, propagates and is destroyed at ``time'' $z^\prime$. The energy
of the ``string'' defect connecting the head to the tail increases
linearly with $|\vec r_\perp -\vec r'_\perp|$ and leads to the exponential
decay of $G(\vec r_\perp, \vec r^\prime_\perp,z,z^\prime)$. Physically, this
represents confinement of the magnetic monopoles represented by the
flux head and tail.
\bigskip
\noindent {\bf Fig.4:}
Two distinct scenarios for vortex crystal melting with increasing
temperature. In type I melting a first order transition separates
a line crystal with $\rho_{\vec G} \neq 0$ from a flux liquid with
$\psi_0 \neq 0$. In type II melting, {\it both} order parameters
are nonzero in the intermediate ``supersolid'' phase.
\bigskip
\noindent {\bf Fig.5:}
Schematic phase diagram of a clean high temperature superconductor.
The ``supersolid'' phase is shown as the shaded region.
In the presence of random pinning, a possible vortex glass transition, at which
the
resistance vanishes, would roughly follow the melting boundary for low fields
and the crystal-supersolid boundary in high fields.
\bigskip
\noindent {\bf Fig.6:}
Initial defect configurations for a symmetric vacancy, centered interstitial,
edge interstitial, split centered and split edge interstitial
used in the numerical calculation of the defect formation energies.
\bigskip
\noindent {\bf Fig.7:}
Starting from the initial symmetric vacancy configuration, shown
as squares, the flux line lattice relaxes first into a symmetric
configuration, shown by the diamonds. This configuration, however,
is just a saddle point and unstable against squeezing the vacancy
along one of the three symmetry axis. One of three degenerate stable final
configurations
is represented by the circles.
\bigskip
\noindent {\bf Fig.8:} Relaxed configurations for centered (squares)
and edge (circle) interstitial. The edge interstitial is unstable with respect
to  a ``buckling'' mode perpendicular to the edge of the triangle. The final
stable configuration is the centered interstitial configuration.
\bigskip
\noindent {\bf Fig.9a:} System size dependence of the formation energies
for centered (solid line), split centred (dot-dashed line) and
edge (dashed line)
interstitials. The energy in units of $E_0$ is plotted
versus the total number of flux lines $N$. The lowest energy configuration
is the centered interstitial.
\bigskip
\noindent {\bf Fig.9b:} System size dependence of the formation energies
for symmetric (solid line), and ``crushed'' (solid line) vacancy. The energy in
units of $E_0$ is plotted versus the total number of flux lines $N$.
The symmetric vacancy configuration (6-fold symmetry)
is unstable against squeezing it along one of the three symmetry axis. The
final stable configuration is the ``crushed'' vacancy with only a two-fold
symmetry.
\bigskip
\noindent {\bf Fig.10:} Relaxed configuration for an edge (open circles)
and split edge (open squares) interstitial. The two cofigurations differ
by a shift along the edge of the triangle. Because of their energy difference
being small gliding along the edge of an triangle is a low energy process.
Note, however, that both configurations are unstable to buckling, resulting in
the centered interstitial configuration.
\bigskip
\noindent {\bf Fig.11a:} Interaction energy of two centered interstitials
in units of $E_0$ versus the distance measured in units of the lattice
constant $a_0$ for two different directions.
(i) The vector connecting the two centered interstitials
points along $\vec a$,  one of the basis vectors of the
unit cell (solid line).
(ii) The vector connecting the two centered interstitials points
along $\vec b$, which is perpendicular to  $\vec a$ (dashed line).
Whereas the interaction
in the a-direction is attractive at small distances it is repulsive in the
perpendicular b-direction. In the b-direction there is a minimum in the
interaction energy at a distance of $d_{\rm min} \approx 2.5 a$.
\bigskip
\noindent {\bf Fig.11b:}
Interaction energy of two ``crushed'' vacancies in units of $E_0$ versus the
distance measured in units of the lattice constant $a$. The vector connecting
the vacancies is pointing {\bf a)} along the two-fold symmetry axis of the
vacancy (solid line) and {\bf b)} perpendicular to it (dashed line).
\bigskip
\noindent {\bf Fig.12:}
Basis vectors $\{\vec\delta_i\}$ for the honeycomb lattice of centered
interstitial sites with lattice constant $b$. The $A$ and $B$ sublattices
are indicated by open and closed circles.
\vfill\eject
\bye